\documentclass[%
a4paper,       
UKenglish,     
DIV10,          
useAMS, referee, doublespace]%
{scrartcl}\usepackage[]{graphicx}\usepackage[]{xcolor}
\makeatletter
\def\maxwidth{ %
  \ifdim\Gin@nat@width>\linewidth
    \linewidth
  \else
    \Gin@nat@width
  \fi
}
\makeatother

\definecolor{fgcolor}{rgb}{0.345, 0.345, 0.345}

\usepackage{framed}
\makeatletter
 {\par\unskip\endMakeFramed%
 \at@end@of@kframe}
\makeatother

\definecolor{shadecolor}{rgb}{.97, .97, .97}
\definecolor{messagecolor}{rgb}{0, 0, 0}
\definecolor{warningcolor}{rgb}{1, 0, 1}
\definecolor{errorcolor}{rgb}{1, 0, 0}
\newenvironment{knitrout}{}{} 

\usepackage{alltt}
\usepackage[]{graphicx}\usepackage[]{color}
\makeatletter
\def\maxwidth{ %
  \ifdim\Gin@nat@width>\linewidth
    \linewidth
  \else
    \Gin@nat@width
  \fi
}
\makeatother
\usepackage[square,numbers,sort&compress]{natbib}
\usepackage{rotating}
\usepackage{alltt}
\usepackage[a4paper, total={6in, 9.25in}]{geometry}
\usepackage{xr-hyper}
\usepackage{authblk}    
\usepackage{hyperref}
\input{header.sty}
\usepackage{color, colortbl}
\usepackage{mathtools}

\newcommand{\alphaoverall}{\alpha_\cap} 

\usepackage{chngcntr}
\usepackage{tikz}
\usetikzlibrary{arrows.meta, shapes, arrows}

\newcolumntype{C}[1]{>{\centering\arraybackslash}p{#1}}
\IfFileExists{upquote.sty}{\usepackage{upquote}}{}

\usepackage[most]{tcolorbox}

\newtcbtheorem{Summary}{\bfseries Summary}{enhanced,drop shadow={black!50!white},
  coltitle=black,
  top=0.3in,
  attach boxed title to top left=
  {xshift=1.5em,yshift=-\tcboxedtitleheight/2},
  boxed title style={size=small,colback=pink}
}{summary}

\newtcolorbox[auto counter]{summary}[1][]{title={\bfseries Box~\thetcbcounter},enhanced,drop shadow={black!50!white},
  coltitle=black,
  top=0.3in,
  attach boxed title to top left=
  {xshift=1.5em,yshift=-\tcboxedtitleheight/2},
  boxed title style={size=small,colback=pink},#1}

\IfFileExists{upquote.sty}{\usepackage{upquote}}{}
\begin{document}

%
\tikzstyle{circle2} = [circle, draw,
    text width=3em, text centered, node distance=3cm]
    
\tikzstyle{block2} = [rectangle, draw,
    text width=11em, text centered, rounded corners, node distance=5.5cm]
\tikzstyle{plain2} = [draw = none, fill = none,
text width=5.5em, minimum height = 1cm]
%
\tikzstyle{plain} = [draw=none, fill=none]

\tikzstyle{line} = [draw, -latex]

\title{Beyond the Two-Trials Rule} 
\ifcase\blinded 
\author{} \or 
\author{Leonhard Held\footnote{Corresponding author,  \texttt{leonhard.held@uzh.ch}}}
\affil{\large University of Zurich\\
Epidemiology, Biostatistics and Prevention Institute (EBPI)\\
  and Center for Reproducible Science (CRS) \\
  Hirschengraben 84, 8001 Zurich, Switzerland}
\date{\large \today}
\fi

\maketitle
\vspace{-.5cm}
\begin{center}
\begin{minipage}{12cm}
  \textbf{Abstract}: 

The two-trials rule for drug approval requires "at least two adequate
and well-controlled studies, each convincing on its own, to establish
effectiveness". This is usually \soutr{employed}\hl{implemented} by requiring two significant
pivotal trials and is the standard regulatory requirement to provide
evidence for a new drug's efficacy. However, there is need to develop
suitable alternatives to this rule for a number of reasons, including the possible
availability of data from more than two trials. 
I consider the case of up to three studies and stress the
importance to control the partial Type-I error rate, where only some
studies have a true null effect, while maintaining the overall Type-I
error rate of the two-trials rule, where all studies have a null
effect.  Some less-known $p$-value combination methods are useful to
achieve this: Pearson's method, Edgington's method and the recently
proposed harmonic mean $\chi^2$-test.  I study their properties and
discuss how they can be extended to a sequential assessment of success
while still ensuring overall Type-I error control.  I compare the
different methods in terms of partial Type-I error rate, project power
and the expected number of studies required.  Edgington's method is
eventually recommended as it is easy to implement and communicate, has
only moderate partial Type-I error rate inflation but substantially
increased project power. 
\end{minipage}
\end{center}

\doublespacing

\clearpage

\section{Introduction}\label{sec:introduction}

The two trials rule is a standard requirement by the FDA to
demonstrate efficacy of drugs. It demands ``at least two pivotal
studies, each convincing on its own''  \citep{FDA1998} before drug approval is
granted \hl{and ``reflects the need for substantiation of experimental results, which has often
been referred to as the need for replication of the finding''} \citep{FDA2019}. The rule is usually implemented by requiring two independent
studies to be significant at the standard (one-sided) $\alpha=0.025$
level \citep[Sec.~12.2.8]{senn:2021}.
Statistical justification for
the two-trials rule is usually based on a hypothesis testing
perspective where Type-I error (T1E) control is the primary goal.  The
T1E rate is the probability of a false claim of success under a certain
null hypothesis. Two
different null hypotheses are relevant if results from two studies are available: 
The
\emph{intersection null hypothesis}
\begin{equation}\label{eq:H0i0}
 H^{\,1}_{0} \cap H^{\,2}_{0}
\end{equation}
is a point null hypothesis, defined as the intersection of the
study-specific null hypotheses $H^{\,i}_{0}$: $\theta_i=0$, $i=1, 2$,
here $\theta_i$ denotes the true effect size in the $i$-th study.  The
probability of a false claim of success with respect to the
intersection null \eqref{eq:H0i0} is the \emph{overall} or
\emph{project-wise} T1E rate \cite{Rosenkranz2023}.  The overall T1E rate of the two-trials
rule is $\alpha^2=0.025^2=0.000625$, because the two studies are
assumed to be independent.  

The \emph{no-replicability} or \emph{union null hypothesis}
is defined as the complement of the
alter\-native hypothesis that the effect is non-null in both studies \citep{Sonnemann1991,Shun_etal2005,Heller_etal2014}. 
This is a composite null hypothesis, which also includes the possibility that
only one of the studies has a null eﬀect:
\begin{equation}\label{eq:H0c0}
 H^{\,1}_{0} \cup H^{\,2}_{0}. 
\end{equation}
I follow \citet{micheloud_etal2023} and call the probability of a false
claim of success with respect to the union null \eqref{eq:H0c0} the
\emph{partial} T1E rate. The partial T1E rate depends on the values of
$\theta_1$ and $\theta_2$, where one of the two parameters must be
zero but the other one may not be zero. The partial T1E rate of the
two-trials rule is bounded from above
by $\alpha$, the exact value depends on the
difference $\theta_1 - \theta_2$ \citep{Zhan_etal2023}.



\hl{The FDA has recently emphasized that ``two positive trials with
  differences in design and conduct may be more persuasive, as
  unrecognized design flaws or biases in study conduct will be less
  likely to impact the outcomes of both trials'' \citep{FDA2019}. 
The FDA also notes that two trials with distinct study populations or
different clinical endpoints provide more evidence of
benefit than two positive identically designed and conducted trials,
which ``could leave the conclusions of both trials vulnerable to any
systematic biases inherent to the particular study design.'' But only in
the latter case the two trials can be considered as fully exchangeable
where pooling the study results with a fixed-effect meta-analysis is a useful alternative
\citep[Sec.~12.2.8]{senn:2021}. However, pooling does not control the
partial T1E rate as one persuasive study may easily overrule the
results from a second unconvincing study.  This
  underlines the importance to control not only the overall but also
  the partial T1E rate.  Of course, all trials considered should be ``adequate and well-controlled'',
  otherwise we may just be interested in the effect estimates from the high quality
  studies \citep{Rubin1992}.
}

\begin{figure}[!ht]
\begin{knitrout}
\definecolor{shadecolor}{rgb}{0.969, 0.969, 0.969}\color{fgcolor}
\includegraphics[width=\maxwidth]{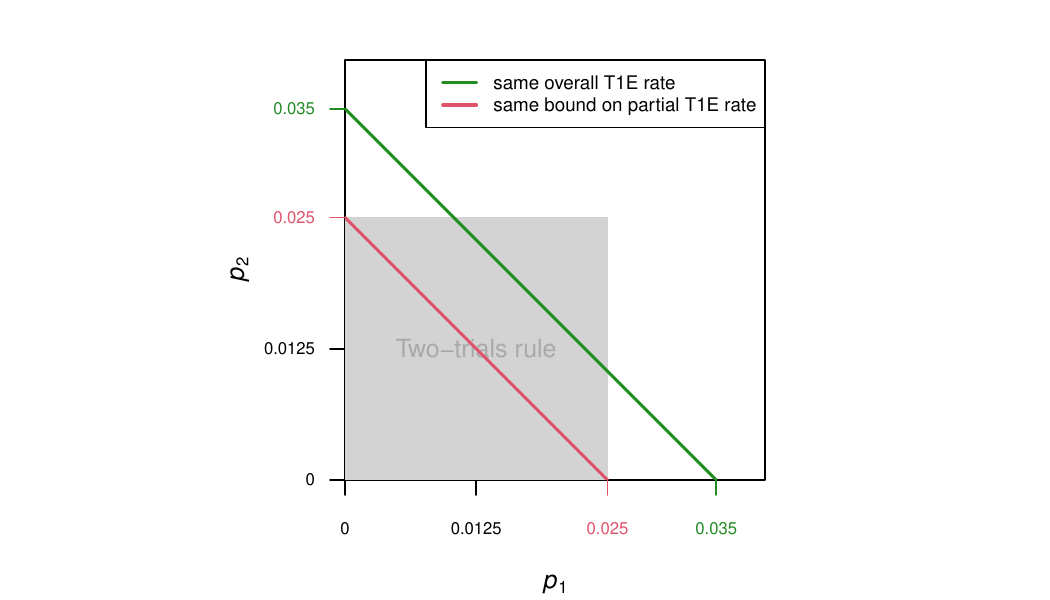} 
\end{knitrout}
\caption{Illustration of partial and overall T1E control with Edgington's method. The grey
  squared region represents the success region of the two-trials rule
  in terms of the two \hl{(one-sided)} $p$-values $p_1, p_2 \leq \alpha = 0.025$, so the partial T1E rate is controlled
  at $\alpha$ and the overall T1E rate is $\alpha^2$. The success region below the
  red line also controls the partial T1E rate at $\alpha$, but has
  reduced overall T1E rate of $\alpha^2/2$. The success region below the green
  line controls the partial T1E rate at $\sqrt{2} \, \alpha \approx 0.035$ and has
overall T1E rate $\alpha^2$, the same as for the two-trials rule. \label{fig:illustration}}
\end{figure}

\hl{In what follows I describe methods that aim to control both the
  overall T1E and partial T1E rates based on results from two (or three) adequately designed trials.} In principle there are two
options to develop alternatives to the two-trials rule, see Figure
\ref{fig:illustration} for an illustration with Edgington's
method\hl{, described in Section \ref{sec:Edgington} in more detail}.  First,
we may consider methods with partial T1E rate bound at $\alpha$, but
this will inevitably reduce the {overall T1E rate} and the project
power, because the success region of any such method (in terms of the
trial-specific $p$-values $p_1$ and $p_2$) must be a subset of the
success region of the two-trials rule. Alternatively we may fix the
overall T1E rate at $\alpha^2$ and allow for some inflation of the
partial T1E rate. Now the success regions is no longer a subset of the
success region of the two-trials rule, so the impact on project power
is not immediate. The latter approach has been also selected by
\citet{Rosenkranz2023} as an ``overarching principle'' in the search
for generalizations of the two-trials rule.  Our goal is thus to allow
for some (limited) inflation of the partial T1E rate while maintaining
overall T1E control.

\hl{Instead of the ``double dichotomization'' of the two-trials rule}
I propose to base inference on $p$-value
combination methods which return a combined
$p$-value that can be interpreted as a quantitative measure of the
total available evidence.
However, many standard \hl{combination} methods do
control the partial T1E rate only at the trivial bound 1, for example
Fisher's or Tippett's
method \cite{Sonnemann1991,cousins2008}.
Also Stouffer's ``inverse normal'' method \citep{Stouffer1949}, which is
closely related to
a fixed effect meta-analysis and called the ``pooled-trials rule'' in
\citet{senn:1997}, does not control the partial T1E rate at a
non-trivial bound. Indeed, all these methods can flag success even if
one of the two studies is completely unconvincing with an effect
estimate perhaps even significant in the wrong direction. This may be
acceptable within a study to be able to stop an experiment at interim
for efficacy \citep{BauerKoehne1994}, but is unacceptable to assess
replicability across studies \citep{Rosenkranz2002}. I therefore
concentrate on less-known $p$-value combination methods that have an
in-built non-trivial control of the partial T1E rate: The $n$-trials
rule \hl{(requiring each of $n$ trials to be significant at a common
  significance level)}, Pearson's and Edgington's method, the harmonic
mean $\chi^2$-test and the sceptical $p$-value.

The \soutr{outcome}\hl{conclusion} of this paper is a tentative recommendation for Edgington's
$p$-value combination method 
\cite{Edgington1972} as summarized in Box \ref{box.edgington}: Declare
success if the sum of the
(one-sided) $p$-values from the two studies is smaller than
$\sqrt{2} \, \alpha \approx 0.035$. A single $p$-value can thus be larger than 0.025 to lead to success (but
not larger than 0.035) as long the other one is
sufficiently small. If three studies are considered, then the sum of
the three $p$-values needs to be smaller than 
$\sqrt[3]{6 \, \alpha^2}\approx 0.16$.
A sequential
version is also described in Box \ref{box.edgington}, where stopping for success after two
studies is possible, otherwise a third study will be required. All
these approaches control the overall T1E rate at $\alpha^2=0.025^2$ while
ensuring that \soutr{all studies}\hl{each study} considered \soutr{are}\hl{is}  ``convincing on its own'' to
a sufficient degree, \ie \hl{Edgington's method has} a non-trivial and sufficiently small bound on the partial T1E rate. Furthermore, the \hl{different versions} have larger power to detect existing
effects than the \hl{two- and three-trials rule, respectively}, and other attractive
properties.

\begin{summary}
  \textbf{Edgington's method: An alternative to the two-trials rule} \\
  Error control: Overall T1E control at level $0.025^2$\\
  Input: One-sided $p$-values $p_1$ and $p_2$ from two trials, possibly $p_3$ from a third trial
  \begin{itemize}
  \item 2 trials in parallel: Flag success if $p_1 + p_2 \leq 0.035$
  \item 3 trials in parallel: Flag success if $p_1 + p_2 + p_3 \leq 0.16$
  \item 3 sequential trials:
  \begin{itemize}
    \item Flag success if after two trials $p_1 + p_2 \leq b_2 = 0.03$
    \item Otherwise flag success if after three trials $p_1 + p_2 + p_3 \leq b_3 = 0.11$
  \end{itemize}
  (Other choices can be made for $b_2$ and $b_3$)
  \end{itemize}
\label{box.edgington}
\end{summary}

Section \ref{sec:combination} describes $p$-value combination
methods with a non-trivial bound on the partial T1E rate.
Section \ref{sec:comparison} compares these methods for data \soutr{from two
respectively three trials}\hl{from two and three trials, respectively}. For three trials we also discuss the 2-of-3
method recently proposed by \citet{Rosenkranz2023}.  Section
\ref{sec:sequential} develops sequential versions of some of the
methods, which allow to stop early for success after two trials.
A comparison in terms of project power and expected number of studies
required is presented. I close with some discussion in Section
\ref{sec:discussion}.

\section{\textit{P}-value combination methods with partial T1E control}\label{sec:combination}
In the following I will describe $p$-value combination methods that
control the partial T1E at a non-trivial bound different from
1. Throughout I will work with one-sided $p$-values $p_1, \ldots, p_n$
and assume that the $p_i$'s are independent and uniformly distributed
under the null hypothesis $H^{\,i}_{0}$: $\theta_i=0$,
$i=1,\ldots,n$. Overall and partial T1E rate are now defined
accordingly across all $n$ trials. 
Throughout I aim to achieve an overall T1E rate rate of $\alphaoverall$, usually $\alphaoverall = 0.025^2 = 0.000625$.

\subsection{The \textit{n}-trials rule}

The two-trials rule can easily be generalized to $n$
trials\hl{, where it is also known as Wilkinson's method \cite{Wilkinson1951}}.
For example, $n=3$ independent trials need to achieve
significance at level $\alpha_\star=\alpha^{2/3}$ to control the
overall T1E rate at level $\alphaoverall = \alpha_\star^3=\alpha^2$. For
$\alpha=0.025$ we obtain
$\alpha_\star=0.085$, so the partial T1E
rate bound of the three-trials rule at overall level $\alphaoverall = 0.025^2$ is
$0.085$.  The general threshold is
$\alpha_\star=\alpha^{2/n}$, which serves as a benchmark for other
methods based on $n$ trials with overall T1E rate $\alphaoverall = \alpha^2$.  
The combined $p$-value of the $n$-trials rule is $p=\max\{p_1, p_2,
\ldots, p_n\}^n.$

\subsection{Pearson's combination test}

Pearson's combination method \citep{Pearson1933,Pearson1934} is a
less-known variation of Fisher's method.  
Fisher's method  \citep{Fisher-1932} is
based on the test statistic
\begin{equation}\label{eq:FP}
F_n = - 2 \sum_{i=1}^n \log(p_i)
\end{equation}
which follows a $\chi^2_{2 n}$ distribution if
the $p$-values $p_1, \ldots, p_n$ are independent uniformly distributed. 
Large values of $F_n$ provide
evidence against the intersection null, and
thresholding $F_n$ at the $(1-\alphaoverall)$-quantile
$\chi^2_{2n}(1-\alphaoverall)$ of the $\chi^2_{2 n}$-distribution gives the success criterion
\begin{equation}\label{eq:fisher}
\prod_{i=1}^n p_i \leq \exp\left\{-0.5 \chi^2_{2n}(1-\alphaoverall) \right\}.
\end{equation}
It follows that a sufficient (but not necessary) criterion for success
is that at least one $p$-value fulfills
\begin{eqnarray}
p_i & \leq & \exp\left\{-0.5 \chi^2_{2n}(1-\alphaoverall) \right\}. \label{eq:FisherBound}
\end{eqnarray}
For example, for $n=2$ and $\alphaoverall=0.025^2=0.000625$, the right-hand
side of \eqref{eq:FisherBound} is
$0.00006$.  If the first $p$-value $p_1$ is
smaller than this bound, Fisher's criterion \eqref{eq:fisher} will flag success
no matter what the result from the second study is and does therefore not control the
partial T1E rate at a non-trivial bound. 

Pearson's method  \citep{Pearson1933,Pearson1934} uses the fact that if
$p_i$ is uniform also $1-p_i$ is uniform, so 
\begin{equation}\label{eq:TP}
K_n = - 2 \sum_{i=1}^n \log(1-p_i)
\end{equation}
also follows a $\chi^2_{2 n}$ distribution under the intersection
null.\footnote{The notation $K$ rather than $P$ for \eqref{eq:TP} is inspired
by Pearson's first name Karl, to avoid confusion with $p$-values
$p$.} Now small values of $K_n$ provide evidence against the
intersection null and we need to threshold $K_n$ at the
$\alphaoverall$-quantile $a_n=\chi^2_{2n}(\alphaoverall)$. This gives
the success criterion $K_n \leq a_n$, or equivalently
\begin{equation}\label{eq:PearsonCriterion}
\prod_{i=1}^n (1-p_i) \geq \exp (-0.5 \, a_n),
\end{equation}
with corresponding combined $p$-value $p=\Pr(\chi^2_{2n} \leq K_n)$.
It follows from \eqref{eq:PearsonCriterion} that a necessary (but not sufficient) success criterion is that
all $p$-values fulfill
\begin{eqnarray}
p_i & \leq & 1 - \exp (-0.5 \, a_n ). \label{eq:PearsonBound}
\end{eqnarray}
For $n=2$ and $\alphaoverall = 0.025^2$ the right-hand
side of \eqref{eq:PearsonBound} 
is $0.035$, which is then
also the bound of Pearson's method on the partial T1E rate and  only slightly larger than for the two-trials rule where both $p$-values have to
be smaller than 0.025.

\subsection{Edgington's method}
\label{sec:Edgington}
\citet{Edgington1972} proposed a method that combines $p$-values by
addition rather than multiplication as in Fisher's criterion \eqref{eq:fisher}.  Under
the intersection null hypothesis, the distribution of the sum of the
$p$-values
\begin{equation}\label{eq:En}
E_n = p_1 + \ldots + p_n
\end{equation}
follows the Irwin-Hall distribution \citep{Irwin1927,Hall1927}, denoted as $E_n \sim \IH(n)$, see
also \citet[Section 26.9]{Johnson.etal1995}. 
A combined $p$-value $p=\Pr(\IH(n) \leq E_n)$ can be calculated using
the cdf of the Irwin-Hall distribution, which is available in closed
form. For $E_n \leq 1$, the computation is particularly simple:
\begin{equation}\label{eq:Ep} 
p = 
  (E_n)^n/n!,
\end{equation}
otherwise a correction term must be added  \citep{Edgington1972}.

Critical values $b_n$, which define success if and only if $E_n \leq b_n$,
can be calculated \soutr{to achieve overall T1E rate $\alphaoverall$}
by replacing $p$ with $\alphaoverall=\alpha^2=0.025^2$ in \eqref{eq:Ep} \hl{and solving for $E_n$}. 
\hl{The definition of $E_n$ in \eqref{eq:En} implies that}
the critical values $b_n$ are also bounds on the partial T1E rate.
For $n=2,3$ we obtain the bounds $b_2=\sqrt{2} \, \alpha=0.035$ and $b_3=\sqrt[3]{6 \alpha^2} = 0.16$, respectively. These bounds can be thought of as a
maximum ``budget'' to be spent on the individual $p$-values to achieve
success. If $n=2$, for example, $p_1 + p_2 \leq
0.035$ is required to flag success at
the $0.025^2$ level, for $n=3$ the success condition is $p_1 + p_2 +
p_3 \leq 0.16$.  The simplicity of
this rule is very attractive in practice, even if more flexibility in
the choice of the partial T1E rate bound may be warranted.

Edgington's method can be viewed as an approximation to Pearson's
method, if all $p$-values are relatively small.
The approximation $\log(1-p_i) \approx -p_i$ then allows to rewrite Pearson's criterion \eqref{eq:PearsonCriterion} to
\begin{equation}\label{eq:PearsonCriterion2}
\sum_{i=1}^n p_i \lesssim  0.5 \chi^2_{2n}(\alphaoverall),
\end{equation}
where the left-hand side is Edgington's test statistic \eqref{eq:En}.
For $n=2$ and
$\alphaoverall=0.025^2$ the
right-hand sided of \eqref{eq:PearsonCriterion2} is
$0.036$, nearly identical to the critical value $b_2=0.035$ based on Edgington's method.

\subsection{Held's method}
The harmonic mean $\chi^2$-test \citep{held2020b}, in the following
abbreviated as Held's method, is a recently proposed 
$p$-value combination method 
that also 
controls the partial T1E rate at a non-trivial bound. 
Consider the inverse normal transformation $Z_i = \Phi(1-p_i)$ of the $p$-values
$p_1, \ldots, p_n$. Under the intersection null hypothesis, all $p$-values are uniform
and the corresponding $Z$-values are therefore independent standard normal. 
The harmonic mean $\chi^2$-test statistic
\[
  X_n^2 = \frac{n^2}{\sum_{i=1}^n 1/Z_i^2} 
\]
then follows a $\chi^2_1$-distribution.  We are
interested in one-sided alternatives where all effect estimates
are positive, say, and flag success if $X_n^2 \geq d_n$ and
 $\sign(Z_1) = \sign(Z_2) = \ldots = \sign(Z_n) = 1$
holds. The critical value 
\begin{equation}\label{eq:alphaH.to.cH}
  d_n = \left[\Phi^{-1}(1-2^{n-1} \alphaoverall) \right]^2
\end{equation}
thus depends on the overall T1E rate
$\alphaoverall$ and the value of $n$ \citep[Section
  2]{held2020b}. \hl{Finally, the combined $p$-value is
  $p =   \Pr(\chi^2(1) \geq X_n^2)/2^n = \left[1-\Phi(X_n)\right]/2^{n-1}$, if all
effect estimates have the anticipated (positive) direction.} 

The requirement $X_n^2 \geq d_n$ is equivalent to 
\begin{equation}\label{eq:Hn}
  H_n = \sum_{i=1}^n 1/Z_i^2 \leq \frac{n^2}{d_n} \eqqcolon c_n.
\end{equation}
From \eqref{eq:Hn} we see that $1/Z_i^2 \leq c_n$ must hold for
all $i=1, \ldots, n$ to achieve success. This can be re-written as 
a necessary (but not sufficient) success condition on the $p$-value from the $i$-th trial: 
\begin{equation}\label{eq:necessary}
p_i \leq 1-\Phi(1/\sqrt{c_n}), \quad i=1, \ldots, n.
\end{equation}
The right-hand side of \eqref{eq:necessary} thus represents a
bound on the partial T1E rate. 
For $\alphaoverall=0.025^2$ and $n=2$ studies we obtain the value
$1 - \Phi(1/\sqrt{0.44}) = 0.065$. This shows that Held's method, applied to two studies, controls the partial T1E
rate, but at a larger bound than Pearson's or Edgington's method.

\subsection{Sceptical \textit{p}-value}

The sceptical $p$-value \citep{held2020} has been developed for the
joint analysis of an original and a replication study, so is
restricted to $n=2$ studies. As the harmonic mean $\chi^2$-test, it
depends on the squared $z$-statistics $Z_1^2$ and $Z_2^2$, but also
takes into account the\soutr{variance} ratio\soutr{$c = \sigma_1^2/\sigma_2^2$, the
ratio} of the variances\soutr{$\sigma_1^2$ and $\sigma_2^2$} of the effect estimates.

\hl{The} originally \hl{proposed} sceptical $p$-value $p_S$ is always
larger than the two study-specific $p$-values, so controls the partial
T1E rate at $\alpha$ if $p_S \leq \alpha$ defines replication
success. Its overall T1E rate is considerably smaller than
$\alpha^2$ and depends on \hl{the variance ratio (original to
replication), in the following denoted by $g$}. \citet{micheloud_etal2023}
 have
recently developed a recalibration that enables exact overall T1E rate
at $\alpha^2$, for any value of the variance ratio\footnote{\hl{denoted by $c$ rather than $g$ in \citet{micheloud_etal2023}}}.
A consequence of this
recalibration is that the bound $\gamma$ on the partial T1E rate is
larger than $\alpha$ and increases with increasing $g$. The limiting
case $g \rightarrow 0$ corresponds to the
two-trials rule. The
harmonic mean $\chi^2$-test for $n=2$ studies turns out to be another
special case for $g=1$. 

Here we propose to use the sceptical $p$-value as a flexible
combination method of $p$-values from two independent trials. The parameter $g$ is then free to
choose and not related to the variance ratio.
It can be selected to
achieve a desired bound $\gamma$ on the partial T1E rate while
maintaining the overall T1E rate at $\alpha^2$, see the inset plot of Figure \ref{fig:fig1}. For example, we might
consider a bound of $\gamma=2 \alpha = 0.05$ on the partial T1E rate
as acceptable \citep{Rosenkranz2002}, in which case
$g=0.43$. The more stringent bound
$\gamma = 1.2 \, \alpha = 0.03$ is obtained with $g=0.04$.

\section{Comparing alternatives to the two-trials rule}\label{sec:comparison}
\subsection{Data from two trials}

\subsubsection{Success regions}
\begin{figure}

\begin{knitrout}
\definecolor{shadecolor}{rgb}{0.969, 0.969, 0.969}\color{fgcolor}
\includegraphics[width=\maxwidth]{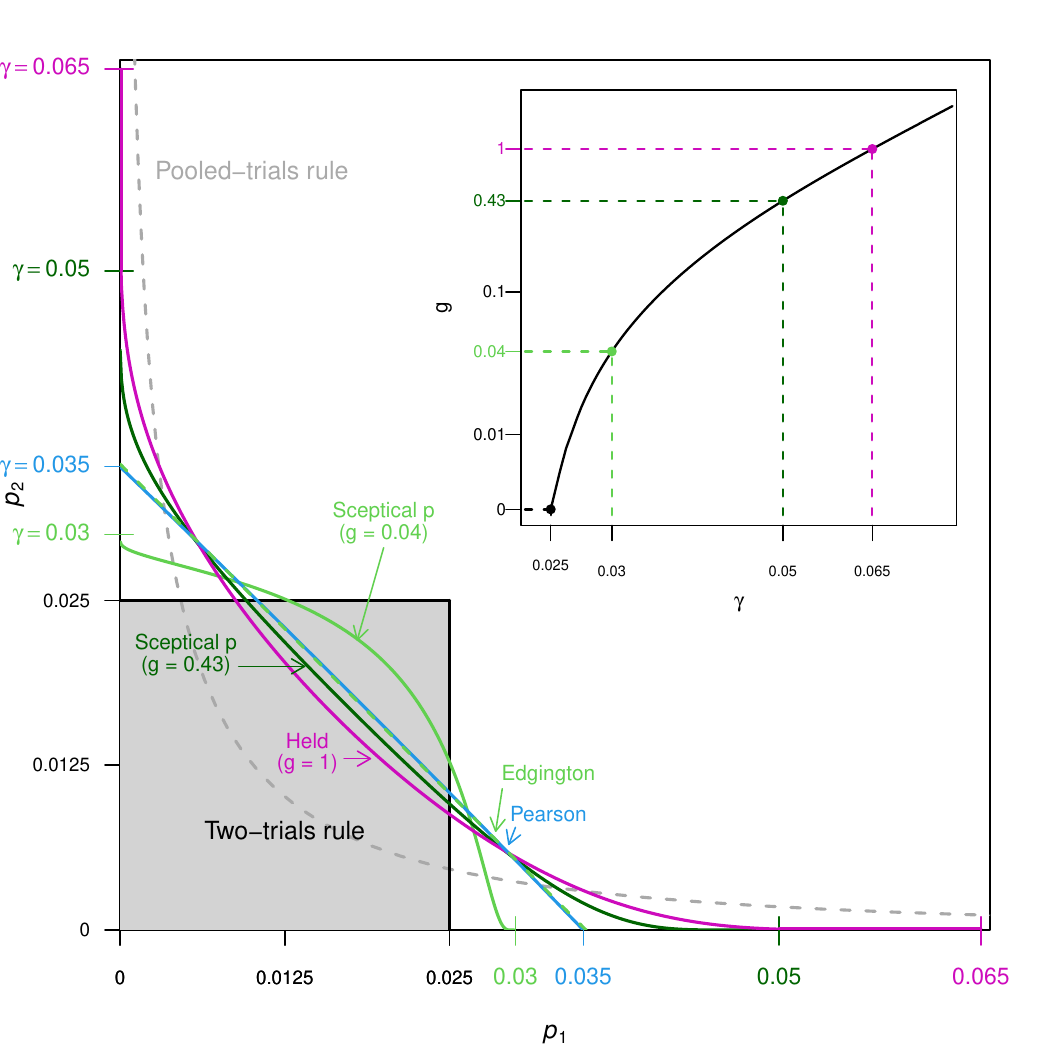} 
\end{knitrout}
\vspace{-0.2cm}
\caption{Success regions of different $p$-value combination methods depending on the two $p$-values $p_1$ and $p_2$. The bound on the partial T1E rate is denoted as $\gamma$. The two-trials rule success region is the squared gray area
  below the black line where $g=0$ and $\gamma = \alpha$.  All methods
  control the overall T1E rate at $\alpha^2=0.025^2\soutr{=
  0.000625}$, the area under each
  curve.  The inset plot shows the dependence of the parameter $g$
  of the sceptical $p$-value on the bound  $\gamma$ of the partial T1E rate.  }
  \label{fig:fig1}
\end{figure}

Figure \ref{fig:fig1} compares the success regions of the two-trials
rule with Pearson's, Edgington's and Held's method for $n=2$ studies, and the sceptical
$p$-value with $g=0.04$ and
$g=0.43$.  The success region of the two-trials rule
requires both $p$-values to be smaller than $\alpha=0.025$ and is
represented by the grey squared area. 
Pearson's method gives a success region nearly a
straight line due to the approximation $\log(1-p_i) \approx -p_i$,
which leads to the approximate Pearson success criterion $p_1 + p_2
\lesssim 0.035$. The success region of
Edgington's method is bounded by an exact straight line and nearly identical.
The success regions of the sceptical $p$-value (with
$g=0.43$) and Held's method are bounded by a convex \soutr{line}\hl{curve} with larger
bounds on the partial T1E rate.  The success region of the sceptical
$p$-value with $g=0.04$ has more overlap with the one
from the two-trials rule.


\hl{Suppose the two $p$-values $p_1$ and $p_2$ originate from two normal
  test statistics $Z_i = \Phi^{-1}(p_i)$, $i=1,2$. Suppose further
  that the two trials have the same size and can be considered as exchangeable, so
  $\theta_1=\theta_2=\theta$. Likelihood theory then implies that
  $Z=Z_1+Z_2$ is sufficient for
  $\theta$. The pooled-trials rule (also shown in Figure
  \ref{fig:fig1}) is based on $Z$, so will then be optimal for
  the test $\theta=0$ vs.~$\theta > 0$, \ie under the intersection null \citep{Maca2002}.
  However, this is no longer true under the union null, where
  $\theta_1 \neq \theta_2$ is possible. As a result, the pooled-trials rule does not control the partial
  T1E rate at a non-trivial bound and success is possible even if one $p$-value is very large. }

\subsubsection{Simulation study}
I now report the success probability of the different
methods under different scenarios. It is well known
\citep{Matthews2006} that under the alternative that was used to power
the two trials, the distribution of $Z_1$ and $Z_2$ is $\Nor(\mu_i,
1)$ where $\mu_i = \Phi^{-1}(1-\alpha) + \Phi^{-1}(1-\beta_i)$, where
$\alpha$ is the significance threshold and $1-\beta_i$ the power of
each trial. The case $\mu_i=0$ (where $\beta_i = 1 - \alpha$)
represents the null hypothesis $H_0^i$: $\theta_i=0$.

We can thus simulate independent $Z_1$ and $Z_2$ for $\alpha=0.025$
and different values of the individual trial power $1-\beta_i \in \{0.9, 0.8, 0.6\}$ and
compute the project power, the proportion of results with drug
approval at overall T1E rate $\alpha^2=0.025^2$.  The simulation is based on \ensuremath{10^{6}}
samples, so the Monte Carlo standard error is smaller than 0.05 on the percentage scale.
The results shown in
Table \ref{tab:tab2} are in the
expected order with increasing project power for increasing bound on
the partial T1E rate.  The increase in project power is \soutr{quite}
substantial compared to the two-trials rule. Already Edgington's method has an increase between 3 and 5 percentage points, the sceptical
$p$-value (with $g=0.43$) increases the project power by 5 to 7
percentage points. Held's method has an even larger
increase, between 6 and 8 percentage points.

We can also investigate the probability of success if one of the true
treatment effects is zero (say $\mu_1=0$), but the other one is not. This
probability cannot be larger than the bound on the partial T1E rate of
the respective method and it is interesting to see how much smaller it is
for different values of the power to detect the effect in the second
study. The results in Table \ref{tab:tab3} show that the probability of a false positive claim has the same ordering as the project power shown in Table \ref{tab:tab2}. The probabilities are fairly small, for example between 1.9 and 3.4\% for the sceptical $p$-value $p_S$ with $g=0.43$ and upper bound 5\%.  Even Held's method with a relatively large bound of 6.5\% has a partial T1E rate of only  3.8\% if the power of the second study is as high as 90\%.

\begin{table}[ht]
\centering
\begingroup\small
\begin{tabular}{rr|rrrrrr}
  \multicolumn{2}{c}{Trial power} &\multicolumn{5}{c}{Method} \\
 \hline
Trial 1 & Trial 2 & Two-trials rule & $p_S(c=0.04)$ & Pearson & Edgington & $p_S(c=0.43)$ & Held \\ 
  \hline
90 & 90 & 81 & 83 & 84 & 84 & 86 & 87 \\ 
  90 & 80 & 72 & 74 & 76 & 76 & 78 & 79 \\ 
  90 & 60 & 54 & 56 & 59 & 59 & 61 & 62 \\ 
   \hline
\end{tabular}
\endgroup
\caption{Project power of different methods for drug approval depending on individual trial power (all entries in \%)} 
\label{tab:tab2}
\end{table}
\begin{table}[ht]
\centering
\begingroup\small
\begin{tabular}{rr|rrrrrr}
  \multicolumn{2}{c}{Trial power} &\multicolumn{5}{c}{Method} \\
 \hline
Trial 1 & Trial 2 & Two-trials rule & $p_S(c=0.04)$ & Pearson & Edgington & $p_S(c=0.43)$ & Held \\ 
  \hline \multicolumn{2}{c}{Upper bound $\gamma$} & 2.5 & 3.0 & 3.5 & 3.5 & 5.0 & 6.5  \\
 \hline
NULL & 90 &  2.2 &  2.5 &  2.9 &  3.0 &  3.4 &  3.8 \\ 
  NULL & 80 &  2.0 &  2.2 &  2.5 &  2.5 &  2.8 &  3.1 \\ 
  NULL & 60 &  1.5 &  1.6 &  1.8 &  1.8 &  1.9 &  2.1 \\ 
   \hline
\end{tabular}
\endgroup
\caption{Partial T1E rate of different methods for drug approval  depending on individual trial power (all entries in \%). NULL indicates a trial with zero effect size. } 
\label{tab:tab3}
\end{table}

\subsection{Data from three trials}

\subsubsection{The 2-of-3 rule}
Additional issues arise in the
application of the two-trials rule if more than two studies are
conducted \citep{Fisher1999b}. Requiring two out of $n$ studies to be
significant at level $\alpha$ inflates the overall T1E rate beyond
$\alpha^2$ and so adjustments are needed.
\citet{Rosenkranz2023} introduces the $k$-of-$n$ rule 
\citep[see also earlier work in][]{Wilkinson1951,ruegger1978}  and
argues that ``the [overall] type-I error rate
of any procedure involving more than two trials shall equal the
[overall] type-I error rate from the two trials rule.''  Here I
consider the ``2-of-3 rule'' where $k=3$ studies are conducted, two of
which need to be significant to flag success.  \citet{Rosenkranz2023} shows that the corresponding
significance level has to be reduced to
$\alpha_{2|3} = 0.0145$ to ensure that the
overall T1E rate is $\alpha^2=0.025^2$. 
However, the 2-of-3 rule no longer controls the partial T1E rate,
because one of the three studies can be completely unconvincing,
perhaps even with an effect size estimate in the wrong direction.
This is a major drawback that may be considered as unacceptable by
regulators. 

Furthermore, thresholding individual studies for significance without
taking into account the results from the other studies creates 
problems in interpretation.  For example, suppose that 
the first two
studies have $p$-values $p_1=p_2=0.02$ while the third study has not
yet been started.  The 2-of-3 rule would then stop the project for
failure, even if the standard two-trials rule (falsely assuming only
two studies have been planned from the start) would flag success. Now
suppose results from the third trial are also available, with
$p_3=0.001$, say.  The 2-of-3 rule would then still conclude project
failure although the combined evidence for an existing treatment
effect 
{would be} overwhelming.  For example, the three-trials rule would flag a clear
success as all three $p$-values are smaller than
$0.085$ and the combined $p$-value $p=\max\{p_1,p_2,p_3\}^3=0.02^3 = 0.000008$ is much smaller than $0.025^2=0.000625$. 
These considerations illustrate that
individual thresholding of study-specific $p$-values may lead to
paradoxes that can be hard to explain to practitioners.

\hl{Nevertheless,  a combined $p$-value can also be calculated for
  the 2-of-3 rule \citep[Chapter 3]{HedgesOlkin1985}, a special case
of Wilkinson's method \citep{Wilkinson1951}.
    First note that success occurs if and only if the
  second smallest $p$-value $p_{(2)}$ is smaller than
  $\alpha_{2|3}$. The second order statistic of three independent
  uniform distributions is known to follow a Beta $\Be(2,2)$ distribution, so the combined $p$-value  of the 2-of-3 rule is $p = \Pr \{\Be(2,2) \leq p_{(2)}\}$.}

\subsubsection{Other methods}\label{sec:other}
I now investigate the applicability of the $p$-value combination
methods described in Section \ref{sec:combination} to data from three trials. 
The sceptical $p$-value is not available for 3 (or more)
studies, so we restrict attention to the remaining methods.
Application of Held's method to $n=3$ studies
requires every study-specific $p$-value $p_i$ to be smaller than
0.175\hl{, as derived from \eqref{eq:necessary}}. This is only slightly larger
than the bounds 0.149 and 0.155 \soutr{with}\hl{from} Pearson's \soutr{respectively}\hl{and} Edgington's
method\hl{, as listed in the last row of Table \ref{tab:tab3b}}.  These values need to be compared not to 0.025, but to
$0.085$, the adjusted significance threshold
of the three-trials rule. The increase in partial T1E rate of Pearson's and Edgington's method is therefore not more than twofold, Held's method has a \hl{slightly larger} increase. \soutr{by a factor of 2.04.}


\subsubsection{Simulation study}
I have conducted another simulation study with $n=3$ trials, powered
at different values, now \hl{each} with significance level
$0.085$. 
The project power listed in Table
\ref{tab:tab2b} shows very similar values for Pearson's, Edgington's
and Held's method. Those values are considerably larger (nearly 10
percentage points) than for the three-trials rule. The 2-of-3 rule is
also listed and has
less increase in project power of
around 3-6 percentage points. 

\begin{table}[ht]
\centering
\begin{tabular}{rrr|rrrrr}
  \multicolumn{3}{c}{Trial power} &\multicolumn{5}{c}{Method} \\
 \hline
Trial 1 & Trial 2 & Trial 3 & Three-trials rule & Pearson & Edgington & Held & 2-of-3 rule \\ 
  \hline
90 & 90 & 90 & 73 & 81 & 81 & 82 & 76 \\ 
  90 & 90 & 80 & 65 & 74 & 74 & 74 & 68 \\ 
  90 & 80 & 60 & 43 & 52 & 53 & 53 & 49 \\ 
   \hline
\end{tabular}
\caption{Project power of different methods for drug approval depending on individual trial power (all entries in \%)} 
\label{tab:tab2b}
\end{table}
\begin{table}[ht]
\centering
\begin{tabular}{rrr|rrrrr}
  \multicolumn{3}{c}{Trial power} &\multicolumn{5}{c}{Method} \\
 \hline
Trial 1 & Trial 2 & Trial 3 & Three-trials rule & Pearson & Edgington & Held & 2-of-3 rule \\ 
  \hline
NULL & 90 & 90 &  6.9 & 10.8 & 11.1 & 11.1 & 46.8 \\ 
  NULL & 90 & 80 &  6.2 &  9.3 &  9.5 &  9.5 & 35.4 \\ 
  NULL & 80 & 60 &  4.1 &  5.7 &  5.8 &  5.8 & 15.4 \\ 
  NULL & NULL & 90 &  0.7 &  0.9 &  0.9 &  1.0 &  2.0 \\ 
  NULL & NULL & 80 &  0.6 &  0.8 &  0.8 &  0.8 &  1.5 \\ 
  NULL & NULL & 60 &  0.4 &  0.5 &  0.5 &  0.6 &  0.8 \\ 
   \hline \multicolumn{3}{c}{Upper bound} & 8.5 & 14.9 & 15.5 & 17.5 & 100  \\
 \hline
\end{tabular}
\caption{Partial T1E rate of different methods for drug approval  depending on individual trial power (all entries in \%).  NULL indicates a trial with true effect size of zero. \hl{The last row gives the upper bounds discussed in Section \ref{sec:other}.}} 
\label{tab:tab3b}
\end{table}

Turning to the partial T1E rates shown in Table \ref{tab:tab3b} we see
values close to 50\% for the 2-of-3 rule if one of the trials comes
from the null and the other two have power 90\% to detect an existing
effect. This illustrates the lack of control of the partial T1E rate.
Pearson's, Edgington's and Held's methods behave again very similar with rates
between 10.8 and 11.1\% \hl{in this extreme scenario.} 
\hl{This is to be compared with  6.9\% partial T1E rate for the three-trials rule.}


\subsubsection{Examples}

We have already considered the example with $p_1=0.02$,
$p_2=0.02$ and $p_3=0.001$, which does not lead to success with the
2-of-3 rule\hl{, which has a combined $p$-value of 0.0012}.
Table \ref{tab:tab3a} \soutr{gives}\hl{compares this to} the combined
$p$-values with the different methods discussed in Section \ref{sec:combination}. All would flag success at
the $\alpha_\cap=0.025^2$ level with the three-trials rule having the
smallest $p$-value\hl{, orders of magnitude smaller than the combined $p$-value of the 2-of-3 rule}. Table \ref{tab:tab3a} also shows
another example with $p_1=0.01$, $p_2=0.01$ and $p_3=0.2$, where the
2-of-3 rule would flag success, 
but all combination
methods would not. Now the three-trials rule has the largest 
combined $p$-value. These examples illustrate that the 2-of-3 rule
behaves very different than the other four methods.

\begin{table}[ht]
\centering
\begin{tabular}{rrr|rrrrr}
  \multicolumn{3}{c}{$P$-values} &\multicolumn{5}{c}{\soutr{Method}\hl{Combined $P$-values}} \\ \hline
Trial 1 & Trial 2 & Trial 3 & Three-trials rule & Pearson & Edgington & Held & \hl{2-of-3 rule} \\ 
  \hline
0.02 & 0.02 & 0.01 & 0.000008 & 0.000021 & 0.000021 & 0.000027 & 0.0012 \\ 
  0.01 & 0.01 & 0.20 & 0.008 & 0.002 & 0.0018 & 0.0031 & 0.0003 \\ 
   \hline
\end{tabular}
\caption{Combined $p$-value with different methods for two examples with $p$-values from three trials.} 
\label{tab:tab3a}
\end{table}

\section{The sequential analysis of up to three studies}\label{sec:sequential}

\begin{sidewaysfigure}
\begin{center}
  \begin{subfigure}[b]{0.27\textwidth}
    \centering
    \resizebox{\linewidth}{!}{
      \begin{tikzpicture}[node distance = 4.5cm, auto]
        
        \node[block2, minimum height = 1.5cm, thick, fill=gray!25] (study1){Trial 1};
        
        \node[block2, minimum height = 1.5cm, thick, fill=gray!25, below of = study1] (study2){Trial 2};
        
        \node[block2, minimum height = 1.5cm, thick, fill=gray!25, below of = study2] (study3){Trial 3};
        
        \node[circle2, minimum height = 1.5cm, thick,fill=red!25,   below = 4cm, left of = study2] (failure2){Failure};
        
        \node[circle2, minimum height = 1.5cm, thick, fill=red!25,  below = 4cm, left of = study3] (failure3){Failure};
        
        \node[circle2, minimum height = 1.5cm, thick,  fill=cyan!25, below = 4cm, right of = study2] (success2){Success};
        
        \node[circle2, minimum height = 1.5cm, thick,  fill=cyan!25, below = 4cm, right of = study3] (success3){Success};

        \path [line] (study1) -- node [midway, right] {}(study2);
        
        \path[line] (study2) -- node[midway, right]{else}(study3);
        
        \path[line] (study2) -- node[midway, left]{$p_1, p_2 > 0.015$}(failure2);
        
        \path[line] (study3) -- node[midway, left]{$p_3 > 0.015$}(failure3);
        
        \path[line] (study2) -- node[midway, right]{$p_1, p_2 \leq 0.015$}(success2);
        
        \path[line] (study3) -- node[midway, right]{$p_3 \leq 0.015$}(success3);
        
      \end{tikzpicture}
    }  
    \caption{\label{fig:1a}{\footnotesize The 2-of-3 rule}}
  \end{subfigure}
  \begin{subfigure}[b]{0.15\textwidth}
  \end{subfigure}
  \begin{subfigure}[b]{0.33\textwidth}
    \centering
    \resizebox{\linewidth}{!}{
      \begin{tikzpicture}[node distance = 4.5cm, auto]

        \node[block2, minimum height = 1.5cm, thick, fill=gray!25] (study1){Trial 1};
        
        \node[block2, minimum height = 1.5cm, thick, fill=gray!25, below of = study1] (study2){Trial 2};
        
        \node[block2, minimum height = 1.5cm, thick, fill=gray!25, below of = study2] (study3){Trial 3};
        
        \node[circle2, minimum height = 1.5cm, thick,fill=red!25,  below = 4cm, left of = study1] (failure1){Failure};
        
        \node[circle2, minimum height = 1.5cm, thick,fill=red!25,   below = 4cm, left of = study2] (failure2){Failure};
        
        \node[circle2, minimum height = 1.5cm, thick, fill=red!25,  below = 4cm, left of = study3] (failure3){Failure};
        
        \node[circle2, minimum height = 1.5cm, thick,  fill=cyan!25, below = 4cm, right of = study3] (success3){Success};

        \path [line] (study1) -- node [midway, right] {else}(study2);
        
        \path[line] (study2) -- node[midway, right]{else}(study3);
        
        \path[line] (study1) -- node[midway, left, align=center]{$p_1 > 0.16$}(failure1);
        
        \path[line] (study2) -- node[midway, left]{$p_1 + p_2 > 0.16$}(failure2);
        
        \path[line] (study3) -- node[midway, left]{$p_1 + p_2 + p_3 > 0.16$}(failure3);
        
        \path[line] (study3) -- node[midway, right]{$p_1 + p_2 + p_3 \leq 0.16$}(success3);

      \end{tikzpicture}
    }
    \caption{\label{fig:1b}{\footnotesize Edgington's method}}
  \end{subfigure}
  \begin{subfigure}[b]{0.33\textwidth}
    \centering
    \resizebox{\linewidth}{!}{
      \begin{tikzpicture}[node distance = 4.5cm, auto]
        
        \node[block2, minimum height = 1.5cm, thick, fill=gray!25] (study1){Trial 1};
        
        \node[block2, minimum height = 1.5cm, thick, fill=gray!25, below of = study1] (study2){Trial 2};
        
        \node[block2, minimum height = 1.5cm, thick, fill=gray!25, below of = study2] (study3){Trial 3};
        
        \node[circle2, minimum height = 1.5cm, thick,fill=red!25,  below = 4cm, left of = study1] (failure1){Failure};
        
        \node[circle2, minimum height = 1.5cm, thick,fill=red!25,   below = 4cm, left of = study2] (failure2){Failure};
        
        \node[circle2, minimum height = 1.5cm, thick, fill=red!25,  below = 4cm, left of = study3] (failure3){Failure};
                
        \node[circle2, minimum height = 1.5cm, thick,  fill=cyan!25, below = 4cm, right of = study2] (success2){Success};
        
        \node[circle2, minimum height = 1.5cm, thick,  fill=cyan!25, below = 4cm, right of = study3] (success3){Success};
        
        \path [line] (study1) -- node [midway, right] {else}(study2);
        
        \path[line] (study2) -- node[midway, right]{else}(study3);
        
        \path[line] (study1) -- node[midway, left, align=center]{$p_1 > 0.11$}(failure1);
        
        \path[line] (study2) -- node[midway, left]{$p_1 + p_2 > 0.11$}(failure2);
        
        \path[line] (study3) -- node[midway, left]{$p_1 + p_2 + p_3  > 0.11$}(failure3);
        
        \path[line] (study2) -- node[midway, right]{$p_1 + p_2 \leq 0.03$}(success2);
        
        \path[line] (study3) -- node[midway, right]{$p_1 + p_2 + p_3 \leq 0.11$}(success3);
        
      \end{tikzpicture}
    }  
    \caption{\label{fig:1c}{\footnotesize Sequential Edgington's method}}
  \end{subfigure}
\end{center}
 \caption{The 2-of-3 rule and Edgington's method to
   analyse three trials sequentially.
   The two-of-three rule requires the adjusted significance level $\alpha_{2|3} = 0.015$ at overall T1E rate $\alphaoverall=0.025^2$.
   Success with Edgington's method  requires $E_3 = \sum_{i=1}^3 p_i$ to be \soutr{smaller} \hl{not larger} than the ``budget'' \soutr{$0.15$}\hl{$0.16$}. Failure occurs as soon as
   $E_k = \sum_{i=1}^k p_i > \soutr{0.15}\hl{0.16}$ for some $k=1,2,3$. The budget is reduced to 0.11 in the sequential application of Edgington's method with $q=0.72$ to be able to stop for success after 2 trials if $p_1 + p_2 \leq 0.03$.
   \label{eq:symbolic1}}
\end{sidewaysfigure}
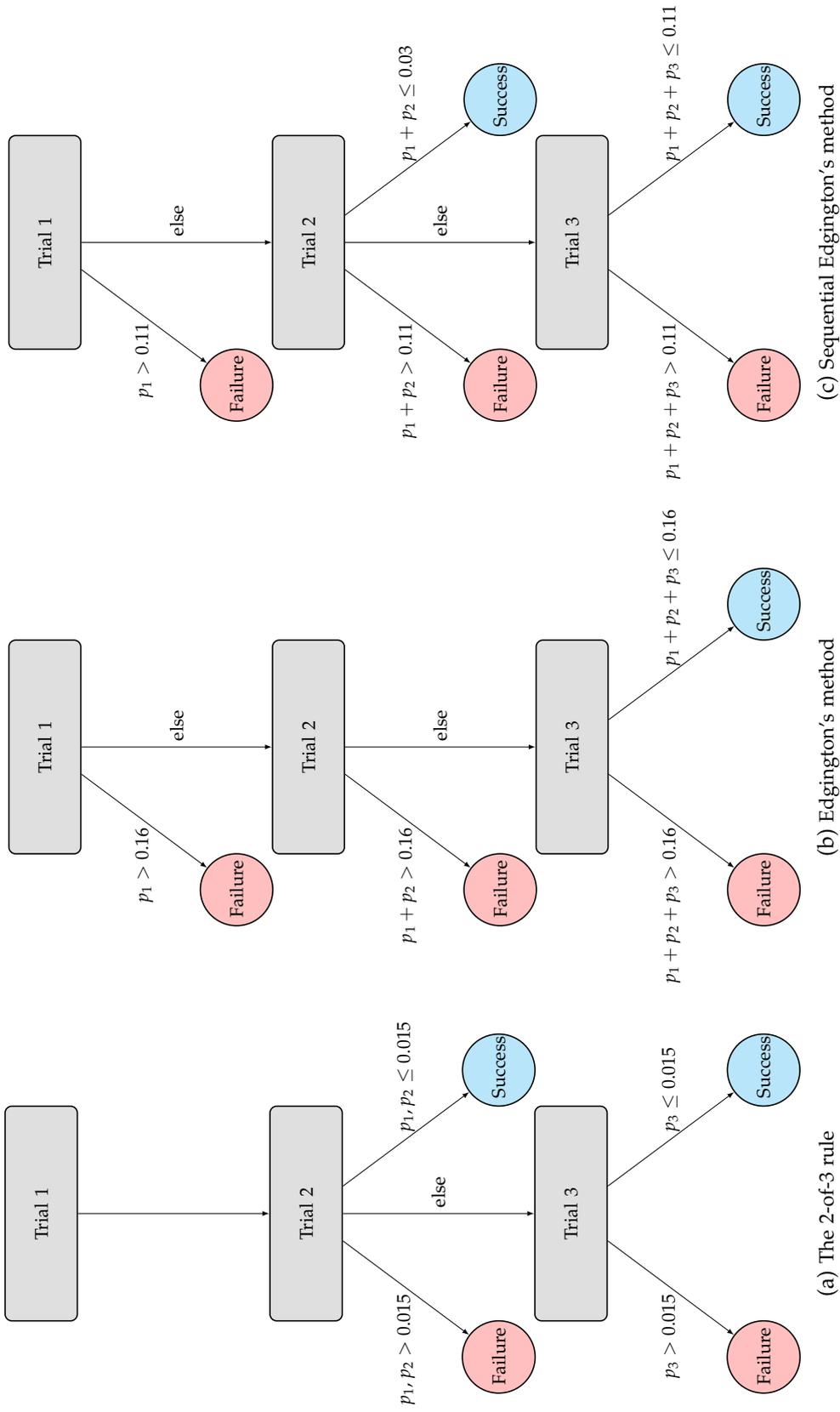

Of particular interest is a sequential application of the different
methods.  The 2-of-3 rule has some advantages here, as it can stop \hl{for efficacy}
already after two studies, see Figure \ref{fig:1a} for
a schematic illustration. Specifically, if the first two studies are
significant at level $\alpha_{2|3}$, a third trial is no longer needed
and resources can be saved. Likewise, if the first two studies are
both not significant, a third trial is pointless because success can
never be achieved. However, there is no mechanism to stop
already after the first trial, even if it is fully
unconvincing.

\subsection{\hl{Early stopping for failure}}
\soutr{In contrast,} All methods that control the partial T1E rate at a
non-trivial bound can stop for failure already after the first and 
second trial, as illustrated 
in Figure \ref{fig:1b} for Edgington's method.  
However, to achieve success these methods  always need three trials. 
This is because they are all based on a ``budget'' $a_3$, $b_3$ resp.~$c_3$ and the results from the
three trials need to be ``within budget'':
\begin{eqnarray*}
  K_3 = \sum_{i=1}^3 - 2 \log(1-p_i) & \leq & a_3 \mbox{ (Pearson), }\\
  E_3 = \sum_{i=1}^3 p_i & \leq & b_3 \mbox{ (Edgington), }\\
  H_3 = \sum_{i=1}^3 1/Z_i^2 & \leq & c_3 \mbox{ (Held). }\\
\end{eqnarray*}

A very convincing trial will cost only very little  and will reduce the
budget only little. On the other hand, an unconvincing first trial is likely to
overspend the budget so success will be impossible, no matter what
the results of the remaining two studies will be. We can hence stop
the project for failure and there is no need to conduct the second and
third study. Likewise, if the budget is overspent after the second
trial, there is no need to conduct the third study.

\begin{figure}

\begin{knitrout}
\definecolor{shadecolor}{rgb}{0.969, 0.969, 0.969}\color{fgcolor}
\includegraphics[width=\maxwidth]{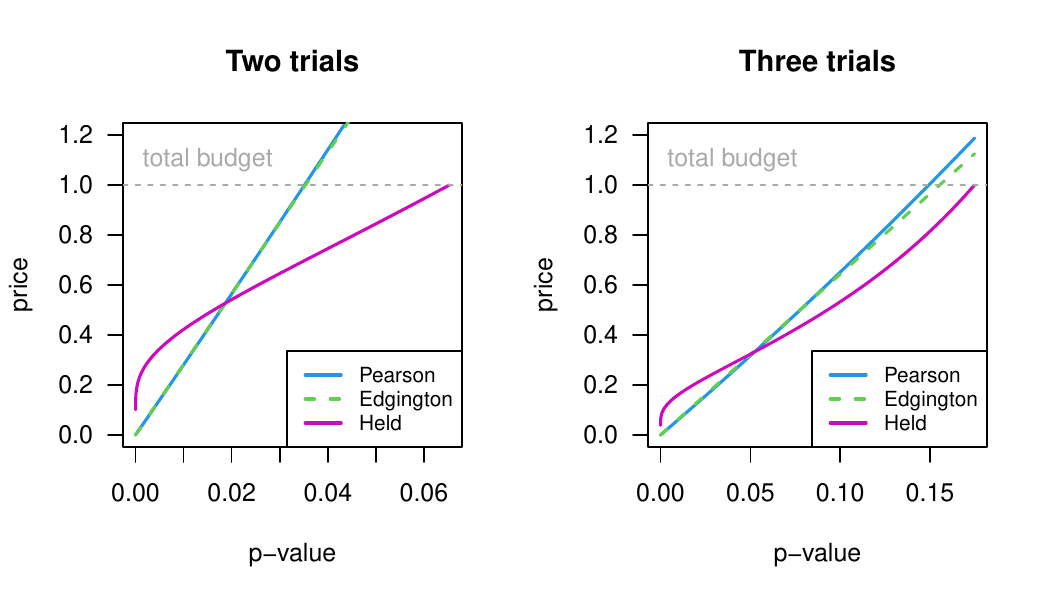} 
\end{knitrout}
\caption{The price of a single study for Pearson's, Edgington's and Held's
  method as a function of the one-sided $p$-value. The price is normalized to a
  unit total budget (the dashed horizontal line) for $n=2$ (left) and $n=3$ (right) trials at overall T1E level $\alphaoverall=0.025^2$.
\label{fig:currency}}
\end{figure}

What is different with the three methods is the ``currency'', either
the ``price'' \hl{of a trial} is given in $-2 \log (1-p_i)$ (Pearson) or $p_i$ (Edgington)
or $1/Z_i^2$ (Held).  This gives different weights for different
$p$-values, as shown in Figure \ref{fig:currency}, where the price of
a single study is normalized to a ``unit budget''. The Figure shows
that the price of Pearson's method is close to Edgington's method, so
nearly linear.  In contrast, Held's method has higher prices of
convincing studies with small $p$-values, while less convincing studies are ``cheaper''. The difference
between Held's and the other methods is larger for $n=2$ than for $n=3$ studies.

\begin{figure}[!h]
\begin{knitrout}
\definecolor{shadecolor}{rgb}{0.969, 0.969, 0.969}\color{fgcolor}
\includegraphics[width=\maxwidth]{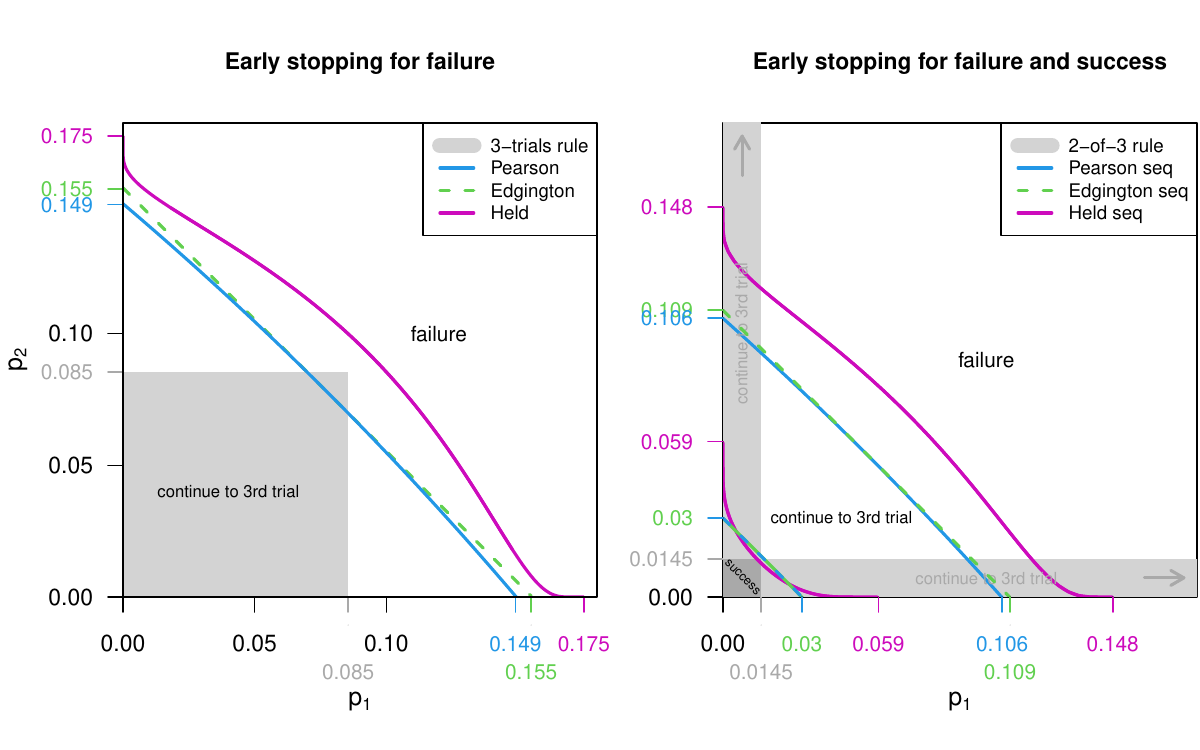} 
\end{knitrout}
\caption{Possible decisions with Pearson's, Edgington's and Held's
  method after two studies with one-sided $p$-values $p_1$ and $p_2$
  have been conducted.  
  The non-sequential procedures (left) will stop for failure if the
  point $(p_1,p_2)$ is above the respective \soutr{solid} line, otherwise a third trial
  will be conducted. The grey areas indicate the corresponding continuation region of the three-trials rule. 
The sequential  procedures (right) are indicated with two lines in the same color and will
  stop for failure if the point $(p_1,p_2)$ is above the upper
  line. They will stop for success if the point is below the lower
  line. If the point is between the upper and lower line, an
  additional third trial will be conducted. The grey areas indicate
  the corresponding success and continuation regions of the 2-of-3 rule.
  \label{fig:decisions2a}}
\end{figure}

Possible decisions after two trials are shown in the left panel of Figure
\ref{fig:decisions2a} for the different methods considered and
compared to the three-trials rule.  Remarkably, the area where the
three-trials rule continues to the third study is considerably smaller
than for the other three methods. Held's method
will always continue with a third study if the three-trials rule does
so, but may also continue if the three-trials rule doesn't.

\subsection{\hl{Early stopping for failure and success}}

Sequential application of one of the methods will also allow to stop for
success after 2 studies.
Adjusted significance levels $\alpha_2$ and $\alpha_3$ for the tests
after \soutr{two respectively three studies}\hl{two and three studies, respectively,} then have to be chosen such that
the overall T1E rate is equal to $\alphaoverall=\alpha^2$.  The level
$\alpha_2 = q \, \alphaoverall$ will be a proportion $q$ of
$\alphaoverall$, following the theory of group-sequential methods
\citep[Section 8.2]{Matthews2006}. Of course, the more lenient we are after the
\soutr{first}\hl{second} study, the more strict we have to be after the \soutr{second}\hl{third}.  I will
choose $q=0.72$ throughout to allow for a 20\% increase of the
partial T1E rate to 0.03 for Edgington's method and
$n=2$ trials, but of course other choices can be made. Sequential
application  with $q=0.72$ is illustrated in
Figure \ref{fig:1c}. As in the non-sequential version, stopping for
failure after Trial 1 or Trial 2 is possible, but now also stopping
for success after two trials.


Computation of the adjusted level $\alpha_3$ to be used after results
from all three trials are available can be done as in group-sequential
trials \citep[Section 8.2.2]{Matthews2006} based on the factorization
\begin{eqnarray}
  K_3 & = & K_2 -2 \log(1-p_3) \label{eq:Pseq} \\
  E_3 & = & E_2 + p_3  \label{eq:Eseq} \\
  H_3 & = & H_2 + 1/Z_3^2  \label{eq:Hseq} 
\end{eqnarray}
for Pearson's, Edgington's and Held's method, respectively. The two terms
on the right-hand side of each equation are independent with known
distributions under the intersection null.  A convolution can be therefore used to compute the
distribution of $K_3$, $E_3$ \soutr{respectively} \hl{and} $H_3$\hl{, respectively,} conditional on no
success after two studies. This can then be used to compute the
adjusted level $\alpha_3$ shown in Table \ref{tab:sequential}, details are given in Appendix \ref{app:sequential}. 
Table \ref{tab:sequential} also gives the corresponding partial T1E bounds $\gamma_2$ resp.~$\gamma_3$ for the sequential methods,
which are smaller than for the non-sequential methods.

\begin{table}
\begin{center}
{
  \begin{tabular}{lrcccc}
    \hline
    & $q$ & $\alpha_2$ & $\alpha_3$ & $\gamma_2$ & $\gamma_3$ \\
    \hline
    Two-trials rule & 1 & $0.025^2$ && 0.025 & \\
    Three-trials rule & 0 & & $0.025^2$ && 0.085 \\ \hline
    Pearson 2 trials & 1 & $0.025^2$ && 0.035 & \\
    Pearson 3 trials & 0 & & $0.025^2$ && 0.15 \\
    Pearson sequential  & 0.72 & $0.0212^2$ & $0.0146^2$ & $0.0298$ & $0.106$ \\ \hline
    Edgington 2 trials & 1 & $0.025^2$ && 0.035 & \\
    Edgington 3 trials & 0 & & $0.025^2$ && 0.16 \\
    Edgington sequential  & 0.72 & $0.0212^2$ & $0.0147^2$ & $0.030$ & $0.109$ \\ \hline
    Held 2 trials & 1 & $0.025^2$ && 0.065 & \\
    Held 3 trials & 0 & & $0.025^2$ && 0.17 \\
    Held sequential  & 0.72 & $0.0212^2$ & $0.0147^2$ & $0.059$ & $0.148$ \\ \hline
  \end{tabular}

}
\end{center}
\caption{Adjusted significance levels $\alpha_i$ and partial T1E
  bounds $\gamma_i$, $i=2,3$, for the two- and three-trials rule, Pearson's,
  Edgington's and Held's method, including sequential versions of the latter three. \label{tab:sequential}}
\end{table}

\subsubsection{Possible decisions after the first trial}

\begin{figure}[!h]

\begin{knitrout}
\definecolor{shadecolor}{rgb}{0.969, 0.969, 0.969}\color{fgcolor}
\includegraphics[width=\maxwidth]{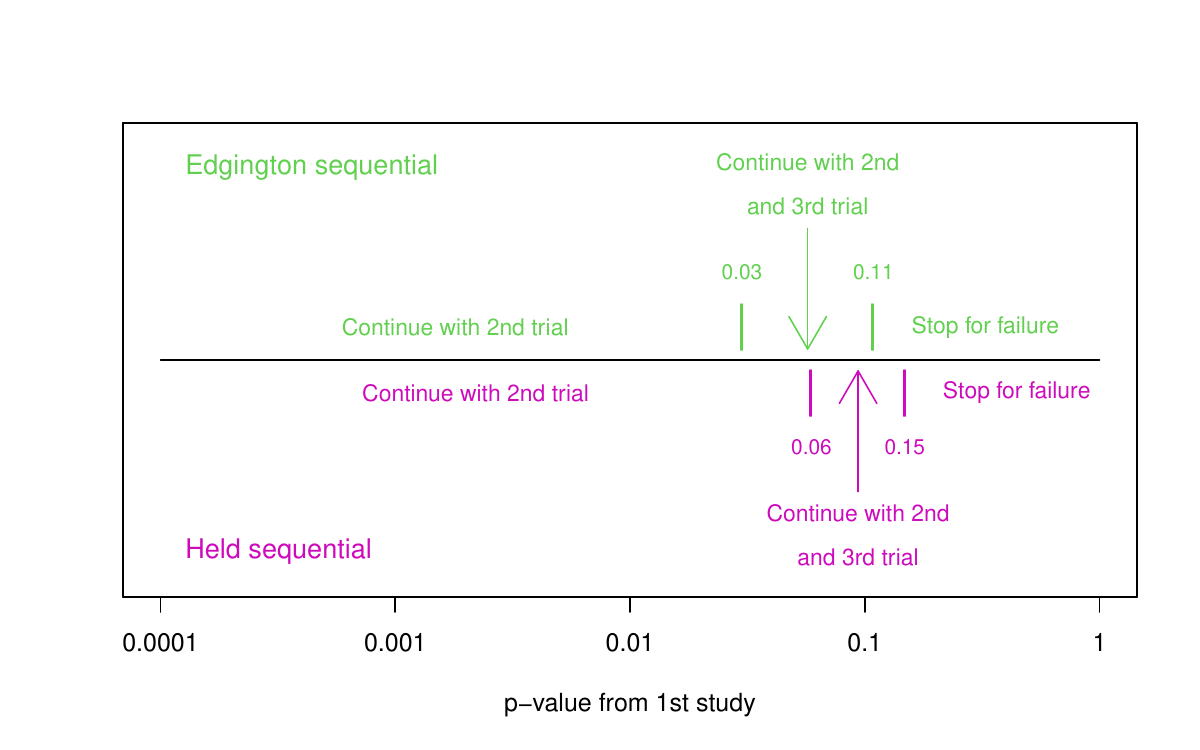} 
\end{knitrout}

\caption{Possible decisions based on the $p$-value $p_1$ from the
  first trial and the sequential Edgington's resp.~Held's method.
  Both methods control the overall T1E rate at
  $\alphaoverall=0.025^2$ and spend the proportion $q=0.72$ of $\alphaoverall$ on $\alpha_2$. \label{fig:decisions1}}
\end{figure}

It is interesting that the sequential versions give three (rather than just two)
options how to proceed after the first study has been conducted.  This is illustrated
in Figure \ref{fig:decisions1} for Edgington's and Held's method and $q=0.72$,
Pearson's method will be very similar to Edgington's method. As we
can see, the three options are stopping for failure if $p_1 >
\gamma_3$, continuing with a second trial (if $p_1 \leq \gamma_2$) or
continuing directly with a second and third trial, perhaps even in
parallel. The last category is chosen, if the $p$-value of the first
study is only ``suggestive'' in the sense that a second trial will
never lead to success but a second and a third trial may do. Formally
this is achieved if $\gamma_2 < p_1 \leq \gamma_3$.

\subsubsection{Possible decisions after the second trial}

The possible decisions of sequential methods (with $q=0.72$) after the
second trial are displayed in the right panel of Figure
\ref{fig:decisions2a}.  Now there is the additional possibility that
the procedures can stop for success. As a result, the failure regions
are somewhat larger than in the non-sequential versions. The 2-of-3
rule is also displayed and has a very different success, continuation
and failure region. \hl{Note that the 2-of-3 continuation region extends up to one, so to
  a larger area than shown, as long as one $p$-value is
  sufficiently small ($<0.0145$).}

\subsection{Operating characteristics}

\begin{figure}
\begin{knitrout}
\definecolor{shadecolor}{rgb}{0.969, 0.969, 0.969}\color{fgcolor}
\includegraphics[width=\maxwidth]{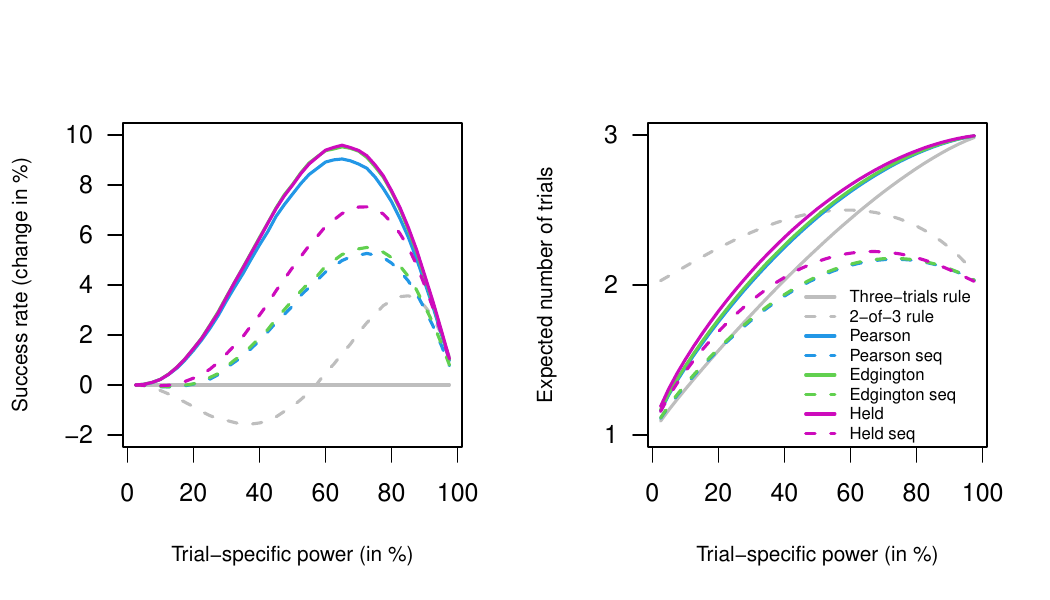} 
\end{knitrout}
\caption{Success rate (left, change in \% compared to the three-trials rule) and number of trials needed (right) for
  the different methods as a function of trial-specific
  power. \label{fig:OC}}
\end{figure}

Figure \ref{fig:OC} compares operating characteristics of the
different methods for data from 3 trials based on a simulation with \ensuremath{10^{7}} samples.
Each individual trial has been powered at the standard significance
level 0.025 with varying power between
2.5\% (where $\mu_i=0$) and 97.5\%. The left
figure gives the difference in success rate compared to the
three-trials rule.  It shows that most of the methods have increased
project power, the largest is obtained for the non-sequential
versions, where Held's and Edgington's method are slightly better than
Pearson's. The sequential version show smaller improvements compared
to the three-trials rule, now with a more pronounced advantage of
Held's method due to the substantially larger bound on the partial T1E rate for 2 trials. 
The 2-of-3 rule has less project power than
the three-trials rule, if the individual trials are underpowered, but more power if the trials
are reasonably powered. 

The right panel of Figure \ref{fig:OC} gives the expected number of
trials required before the project is stopped. If the power is low,
the 2-of-3 rule has the largest expected number of studies
required. All other methods can stop for failure already after trial
1, so the expected number of trials is between 1 and 2 for underpowered
studies. If the power is large, the non-sequential methods require the
largest number of studies on average, because they require three studies to be conducted to reach success. 
The sequential methods of Pearson's, Edgington's and Held's method
require
a smaller number of studies, never larger than around 2.2 studies on average.
\hl{Of course, all these results are based on a fixed proportion $q=0.72$ of overall T1E rate spent 
  after the second study. It remains to be investigated
  which choice of $q$ gives optimal operating characteristics
  in terms of project power or expected number of trials.}

\section{Discussion and Outlook}\label{sec:discussion}

Alternatives to the two-trials rule require appropriate T1E control,
both overall and partial.  I have described and compared different
$p$-value combination methods that have the same overall T1E rate,
offer partial T1E control at a non-trivial bound and can be extended
to a sequential assessment of success. The methods have different
properties and a particular one can be chosen based on a desired
bound on the partial T1E rate. Edgington's method based on the sum of
the $p$-values has the advantage that it is easy to implement and
communicate, has only moderate partial Type-I error rate inflation but
still substantially increased project power compared to the two-
\soutr{respectively}\hl{and} three-trials rule\hl{, respectively}.
Of course, to guarantee exact overall T1E rate control, the specific
type of combination method, whether fixed or sequential as well as the
maximum number of studies conducted have to be defined in advance of
the project \hl{in a ``master protocol''}.  This may be considered a limitation from a study conduct
point of view \citep{Rosenkranz2023}.

\hl{A related question is how acceptable it is by regulators to have inflated partial T1E.
\citet{Johnston_etal2023} systematically investigated the frequency of and rationale for FDA approval of drugs based on pivotal trials with non-significant (``null'') findings. Among 210 new drugs approved between 2018 and 2021  they identified 21 (10\%) not meeting pivotal trial endpoints. 15 were based on more than 1 pivotal study and the most common reported rationale for FDA approval was ``success of other primary end points in pivotal study'' (13 drugs). This implies that regulators do implicitly allow for some partial T1E rate inflation as long as there is another successful (i.e.~conventionally significant) trial. The proposed methods in this paper formalize this rationale for drug approval, requiring overall T1E control at $\alpha^2$ while allowing for a partial T1E rate somewhat larger than $\alpha$. }

\soutr{The sequential versions of the different combination methods have the
advantage that they can stop for success already after two
trials. For $q=0.72$, the probability that a third study has to be conducted
is bounded by 30\% (Pearson),
30\% (Edgington) \soutr{respectively} \hl{and}
34\% (Held), \hl{respectively,} assuming that all
three trials have the same power to detect the true effect, see Figure
\ref{fig:Stop} in Appendix \ref{app:stopping}. 
It remains to be investigated
which choice of $q$ gives optimal operating characteristics, for
example in terms of project power or the expected number of trials.}

In the simulations presented I have assumed that the sample size of
each trial was calculated in advance based on a \hl{certain power
  $1-\beta$ to detect a} pre-specified clinically relevant difference.
\hl{It was shown that the methods described have larger project power
  than the two- and three trials rule, respectively.  One may also
  ask how much the sample size of two trials can be
  reduced to achieve the project power $(1-\beta)^2$ of the two-trials
  rule. The possible reduction in sample size is around 6\% for
  Edgington's rule (see Appendix \ref{app:SSReduction}), for the other
  methods even larger.
}

The different methods also
allow to compute the sample size adaptively based on the results from
the previous trial \citep{Deforth_etal2023}. For example,
we could design a second trial to detect the
observed effect from the first trial with a certain power \citep{MicheloudHeld2022}. This will
be particularly simple for Edgington's method, where the required
significance level is $b_2-p_1$, but also straightforward for the
other approaches.  A comparison of the expected number of patients
needed under the different procedures could be done and the
conditional T1E rate could be investigated \citep[Section
  3.4]{micheloud_etal2023}. We may also compute the conditional (or
predictive) power to reach success given the results from an initial
trial. The bound on the partial T1E rate can directly be used to
describe when this power is exactly zero and when not. But even if
conditional power is non-zero it can be too small to warrant a further
trial. \soutr{Conditional power can be computed directly for $n=2$ from the
results described in \citet{micheloud_etal2023}, otherwise Monte Carlo
simulation \soutr{seems straightforward} \hl{is still possible}.}

It is also possible to adopt the
sequential methods such that a stop for success already after the
first study would be possible. This would incorporate the argument
brought forward by \citet{Fisher1999} that one large and very
convincing trial may give sufficient evidence of efficacy. \soutr{, but would
have to be very convincing.} Also Kay \cite{kay:2015} notes that
``where there are practical reasons why two trials cannot be easily
undertaken or where there is a major unfulfilled public health need,
it may be possible for a claim to be based on a single pivotal
trial.'' The notion of replicability could then be enforced in a
post-marketing requirement after
conditional or accelerated drug approval \citep{Deforth_etal2023}.
However, it is well-known from group-sequential methods, that the $p$-value
obtained after stopping for success overstates the evidence against
the null and some adjustment is required \citep{Cook2002}. 

\hl{Finally, regulators are also interested in a combined effect
  estimate with confidence interval to assess whether the project has
  demonstrated a clinically meaningful effect \citep{FDA2019}.}  $P$-value combination
methods can be inverted to obtain a confidence
interval for the effect estimate based on the corresponding $p$-value
function \citep{Infanger2019}. In future work we will compare the
different methods, for example in terms of coverage and width of the
resulting confidence intervals. We will also investigate how they
need to be adopted to the sequential setting, where the standard
combined effect estimate in group-sequential trials is known to be
biased if we stop early for success \citep{PinheiroDeMets1997}.

\paragraph*{Code}
Code to reproduce the calculations in this manuscript is available at \url{https://osf.io/3pgmk/}.

\paragraph*{Acknowledgments}

Support by the Swiss National Science Foundation (Project \# 189295)
is gratefully acknowledged. I appreciate helpful comments by Charlotte
Micheloud, Samuel Pawel, Frank P\'{e}tavy and Kit Roes. \hl{I am
  grateful for comments by two referees on an earlier version of this
  paper.}  

 \bibliographystyle{unsrtnat}
\bibliography{antritt}

\begin{thebibliography}{39}
\providecommand{\natexlab}[1]{#1}
\providecommand{\url}[1]{\texttt{#1}}
\expandafter\ifx\csname urlstyle\endcsname\relax
  \providecommand{\doi}[1]{doi: #1}\else
  \providecommand{\doi}{doi: \begingroup \urlstyle{rm}\Url}\fi

\bibitem[FDA(1998)]{FDA1998}
FDA.
\newblock Providing clinical evidence of effectiveness for human drug and
  biological products, 1998.
\newblock
  \url{www.fda.gov/regulatory-information/search-fda-guidance-documents/providing-clinical-evidence-effectiveness-human-drug-and-biological-products}.

\bibitem[FDA(2019)]{FDA2019}
FDA.
\newblock Substantial evidence of effectiveness for human drug and biological
  products, 2019.
\newblock
  \url{https://www.fda.gov/drugs/guidance-compliance-regulatory-information/guidances-drugs}.

\bibitem[Senn(2021)]{senn:2021}
S.~Senn.
\newblock \emph{Statistical Issues in Drug Development}.
\newblock John Wiley \& Sons, Chichester, U.K., third edition, 2021.

\bibitem[Rosenkranz(2023)]{Rosenkranz2023}
G.~K. Rosenkranz.
\newblock A generalization of the two trials paradigm.
\newblock \emph{Therapeutic Innovation \& Regulatory Science}, 57:\penalty0
  316--320, 2023.
\newblock URL \url{https://doi.org/10.1007/s43441-022-00471-4}.

\bibitem[Sonnemann(1991)]{Sonnemann1991}
E.~Sonnemann.
\newblock {Kombination unabhängiger Tests}.
\newblock In Jürgen Vollmar, editor, \emph{Biometrie in der
  chemisch-pharmazeutischen Industrie 4}, pages 91--112. Fischer Verlag, 1991.
\newblock in German.

\bibitem[Shun et~al.(2005)Shun, Chi, Durrleman, and Fisher]{Shun_etal2005}
Z.~Shun, E.~Chi, S.~Durrleman, and L.~Fisher.
\newblock Statistical consideration of the strategy for demonstrating clinical
  evidence of effectiveness—one larger vs two smaller pivotal studies.
\newblock \emph{Statistics in Medicine}, 24\penalty0 (11):\penalty0 1619--1637,
  2005.
\newblock \doi{10.1002/sim.2018}.
\newblock URL \url{https://onlinelibrary.wiley.com/doi/abs/10.1002/sim.2018}.

\bibitem[Heller et~al.(2014)Heller, Bogomolov, and Benjamini]{Heller_etal2014}
R.~Heller, M.~Bogomolov, and Y.~Benjamini.
\newblock Deciding whether follow-up studies have replicated findings in a
  preliminary large-scale omics study.
\newblock \emph{Proceedings of the National Academy of Sciences}, 111\penalty0
  (46):\penalty0 16262--16267, 2014.
\newblock \doi{10.1073/pnas.1314814111}.
\newblock URL \url{https://www.pnas.org/doi/abs/10.1073/pnas.1314814111}.

\bibitem[Micheloud et~al.(2023)Micheloud, Balabdaoui, and
  Held]{micheloud_etal2023}
C.~Micheloud, F.~Balabdaoui, and L.~Held.
\newblock Assessing replicability with the sceptical $p$-value: Type-{I} error
  control and sample size planning.
\newblock \emph{Statistica Neerlandica}, 77:\penalty0 573--591, 2023.
\newblock \url{https://doi.org/10.1111/stan.12312}.

\bibitem[Zhan et~al.(2023)Zhan, Kunz, and Stallard]{Zhan_etal2023}
S.~J. Zhan, C.~U. Kunz, and N.~Stallard.
\newblock Should the two-trial paradigm still be the gold standard in drug
  assessment?
\newblock \emph{Pharmaceutical Statistics}, 22:\penalty0 96--111, 2023.
\newblock \doi{https://doi.org/10.1002/pst.2262}.
\newblock URL \url{https://onlinelibrary.wiley.com/doi/abs/10.1002/pst.2262}.

\bibitem[Rubin(1992)]{Rubin1992}
D.~B. Rubin.
\newblock {Meta-analysis: literature synthesis or effect-size surface
  estimation?}
\newblock \emph{Journal of Educational Statistics}, 17\penalty0 (4):\penalty0
  363--374, 1992.

\bibitem[Cousins(2008)]{cousins2008}
Robert~D. Cousins.
\newblock Annotated bibliography of some papers on combining significances or
  p-values, 2008.
\newblock \url{https://arxiv.org/abs/0705.2209}.

\bibitem[Stouffer et~al.(1949)Stouffer, Suchman, Devinney, Star, and
  Williams]{Stouffer1949}
S.~A. Stouffer, E.~A. Suchman, L.~C. Devinney, S.~A. Star, and Jr. Williams,
  R.~M.
\newblock \emph{The American soldier: Adjustment during army life. (Studies in
  social psychology in World War II)}.
\newblock Cambridge University Press, Princeton Univ. Press, 1949.

\bibitem[Senn(1997)]{senn:1997}
S.~Senn.
\newblock \emph{Statistical Issues in Drug Development}.
\newblock John Wiley \& Sons, Chichester, U.K., first edition, 1997.

\bibitem[Bauer and K\"ohne(1994)]{BauerKoehne1994}
P.~Bauer and K.~K\"ohne.
\newblock Evaluation of experiments with adaptive interim analyses.
\newblock \emph{Biometrics}, 50:\penalty0 1029--1041, 1994.
\newblock URL \url{http://www.jstor.org/stable/2533441}.

\bibitem[Rosenkranz(2002)]{Rosenkranz2002}
G.~Rosenkranz.
\newblock Is it possible to claim efficacy if one of two trials is significant
  while the other just shows a trend?
\newblock \emph{Drug Information Journal}, 36\penalty0 (1):\penalty0 875--879,
  2002.
\newblock URL \url{https://doi.org/10.1177/009286150203600416}.

\bibitem[Edgington(1972)]{Edgington1972}
E.~S. Edgington.
\newblock An additive method for combining probability values from independent
  experiments.
\newblock \emph{The Journal of Psychology}, 80:\penalty0 351--363, 1972.

\bibitem[Wilkinson(1951)]{Wilkinson1951}
B.~Wilkinson.
\newblock A statistical consideration in psychological research.
\newblock \emph{Psychological Bulletin}, 48:\penalty0 156--157, 1951.
\newblock \doi{10.1037/h0059111}.

\bibitem[Pearson(1933)]{Pearson1933}
K.~Pearson.
\newblock On a method of determining whether a sample of size n supposed to
  have been drawn from a parent population having a known probability integral
  has probably been drawn at random.
\newblock \emph{Biometrika}, 25:\penalty0 379--410, 1933.
\newblock URL \url{http://www.jstor.org/stable/2332290}.

\bibitem[Pearson(1934)]{Pearson1934}
K.~Pearson.
\newblock On a new method of determining "goodness of fit".
\newblock \emph{Biometrika}, 26:\penalty0 425--442, 1934.
\newblock URL \url{http://www.jstor.org/stable/2331988}.

\bibitem[Fisher(1932)]{Fisher-1932}
R.~A. Fisher.
\newblock \emph{{Statistical Methods for Research Workers}}.
\newblock Oliver \& Boyd, Edinburgh, 4th ed. edition, 1932.

\bibitem[Irwin(1927)]{Irwin1927}
J.~O. Irwin.
\newblock {On the frequency distribution of the means of samples from a
  population having any law of frequency with finite moments, with special
  reference to Pearson's Type II}.
\newblock \emph{Biometrika}, 19\penalty0 (3-4):\penalty0 225--239, 12 1927.
\newblock ISSN 0006-3444.
\newblock \doi{10.1093/biomet/19.3-4.225}.
\newblock URL \url{https://doi.org/10.1093/biomet/19.3-4.225}.

\bibitem[Hall(1927)]{Hall1927}
P.~Hall.
\newblock {The distribution of means for samples if size $n$ drawn from a
  population in which the variate takes values between 0 and 1, all such values
  being equally probable}.
\newblock \emph{Biometrika}, 19\penalty0 (3-4):\penalty0 240--244, 12 1927.
\newblock ISSN 0006-3444.
\newblock \doi{10.1093/biomet/19.3-4.240}.
\newblock URL \url{https://doi.org/10.1093/biomet/19.3-4.240}.

\bibitem[Johnson et~al.(1995)Johnson, Kotz, and Balakrishnan]{Johnson.etal1995}
N.~L. Johnson, S.~Kotz, and N.~Balakrishnan.
\newblock \emph{{C}ontinuous {U}nivariate {D}istributions, Volume 2}, volume~2.
\newblock Wiley, New York, 2 edition, 1995.

\bibitem[Held(2020{\natexlab{a}})]{held2020b}
L.~Held.
\newblock The harmonic mean $\chi^2$-test to substantiate scientific findings.
\newblock \emph{Journal of the Royal Statistical Society, Series C},
  69\penalty0 (3):\penalty0 697--708, 2020{\natexlab{a}}.

\bibitem[Held(2020{\natexlab{b}})]{held2020}
L.~Held.
\newblock A~new standard for the analysis and design of replication studies
  (with discussion).
\newblock \emph{Journal of the Royal Statistical Society, Series A},
  {183}:\penalty0 431--469, 2020{\natexlab{b}}.

\bibitem[Maca et~al.(2002)Maca, Gallo, Branson, and Maurer]{Maca2002}
J.~Maca, P.~Gallo, M.~Branson, and W.~Maurer.
\newblock Reconsidering some aspects of the two-trials paradigm.
\newblock \emph{Journal of Biopharmaceutical Statistics}, 12\penalty0
  (2):\penalty0 107--119, jan 2002.
\newblock \doi{10.1081/bip-120006450}.
\newblock URL \url{https://doi.org/10.1081/bip-120006450}.

\bibitem[Matthews(2006)]{Matthews2006}
J.~N.~S. Matthews.
\newblock \emph{{I}ntroduction to {R}andomized {C}ontrolled {C}linical
  {T}rials}.
\newblock Chapman \& Hall/CRC, second edition, 2006.

\bibitem[Fisher(1999{\natexlab{a}})]{Fisher1999b}
L.~D. Fisher.
\newblock Carvedilol and the {Food and Drug Administration} ({FDA}) approval
  process: The {FDA} paradigm and reflections on hypothesis testing.
\newblock \emph{Controlled Clinical Trials}, 20\penalty0 (1):\penalty0 16--39,
  1999{\natexlab{a}}.
\newblock ISSN 0197-2456.
\newblock \doi{https://doi.org/10.1016/S0197-2456(98)00054-3}.
\newblock URL
  \url{http://www.sciencedirect.com/science/article/pii/S0197245698000543}.

\bibitem[Rüger(1978)]{ruegger1978}
B.~Rüger.
\newblock {Das maximale Signifikanzniveau des Tests: "Lehne $H_0$ ab, wenn $k$
  unter $n$ gegebenen Tests zur Ablehnung führen".}
\newblock \emph{Metrika}, 25:\penalty0 171--178, 1978.

\bibitem[Hedges and Olkin(1985)]{HedgesOlkin1985}
Larry~V. Hedges and Ingram Olkin.
\newblock \emph{Statistical Methods for Meta-Analysis}.
\newblock Elsevier, 1985.
\newblock \doi{10.1016/c2009-0-03396-0}.

\bibitem[Johnston et~al.(2023)Johnston, Ross, and
  Ramachandran]{Johnston_etal2023}
J.~L. Johnston, J.~S. Ross, and R.~Ramachandran.
\newblock {US Food and Drug Administration Approval of Drugs Not Meeting
  Pivotal Trial Primary End Points, 2018-2021}.
\newblock \emph{JAMA Internal Medicine}, 183\penalty0 (4):\penalty0 376--380,
  04 2023.
\newblock ISSN 2168-6106.
\newblock \doi{10.1001/jamainternmed.2022.6444}.
\newblock URL \url{https://doi.org/10.1001/jamainternmed.2022.6444}.

\bibitem[Deforth et~al.(2023)Deforth, Micheloud, Roes, and
  Held]{Deforth_etal2023}
M.~Deforth, C.~Micheloud, K.~Roes, and L.~Held.
\newblock Combining evidence from clinical trials in conditional or accelerated
  approval.
\newblock \emph{Pharmaceutical Statistics}, 22:\penalty0 707--720, 2023.
\newblock \url{https://doi.org/10.1002/pst.2302}.

\bibitem[Micheloud and Held(2022)]{MicheloudHeld2022}
C~Micheloud and L~Held.
\newblock {Power calculations for replication studies}.
\newblock \emph{Statistical Science}, 37\penalty0 (3):\penalty0 369--379, 2022.
\newblock \doi{10.1214/21-STS828}.
\newblock URL \url{https://doi.org/10.1214/21-STS828}.

\bibitem[Fisher(1999{\natexlab{b}})]{Fisher1999}
L.~D. Fisher.
\newblock One large, well-designed, multicenter study as an alternative to the
  usual {FDA} paradigm.
\newblock \emph{Drug Information Journal}, 33\penalty0 (1):\penalty0
  265--–271, 1999{\natexlab{b}}.
\newblock URL \url{https://doi.org/10.1177/009286159903300130}.

\bibitem[Kay(2015)]{kay:2015}
R.~Kay.
\newblock \emph{Statistical Thinking for Non-Statisticians in Drug Regulation}.
\newblock John Wiley \& Sons, Chichester, U.K., second edition, 2015.
\newblock \url{https://doi.org/10.1002/9781118451885}.

\bibitem[Cook(2002)]{Cook2002}
T.~D. Cook.
\newblock P-value adjustment in sequential clinical trials.
\newblock \emph{Biometrics}, 58:\penalty0 105--11, 2002.
\newblock URL \url{http://www.jstor.org/stable/3068544}.

\bibitem[Infanger and Schmidt-Trucksäss(2019)]{Infanger2019}
D.~Infanger and A.~Schmidt-Trucksäss.
\newblock P value functions: An underused method to present research results
  and to promote quantitative reasoning.
\newblock \emph{Statistics in Medicine}, 38\penalty0 (21):\penalty0 4189--4197,
  2019.
\newblock \doi{10.1002/sim.8293}.
\newblock URL \url{https://onlinelibrary.wiley.com/doi/abs/10.1002/sim.8293}.

\bibitem[Pinheiro and DeMets(1997)]{PinheiroDeMets1997}
J.~C. Pinheiro and D.~L. DeMets.
\newblock Estimating and reducing bias in group sequential designs with
  {G}aussian independent increment structure.
\newblock \emph{Biometrika}, 84:\penalty0 831--845, 1997.

\bibitem[Hung et~al.(1997)Hung, O'Neill, Bauer, and K\"ohne]{Hung_etal1997}
H~M~J Hung, R~T O'Neill, P~Bauer, and K~K\"ohne.
\newblock The behavior of the {$P$}-value when the alternative hypothesis is
  true.
\newblock \emph{Biometrics}, 53\penalty0 (1):\penalty0 11--22, 1997.

\end{thebibliography}

\begin{appendix}

\section{Details on sequential application}\label{app:sequential}

\subsection{Pearson's method}

\hl{The adjusted level $\alpha_2 = q \cdot \alpha_\cap$ with budget
$a_2=\chi^2_4(\alpha_2)$ is chosen to spend the proportion $q$ of $\alpha_\cap$ on $\alpha_2$.}
To compute \hl{the adjusted level $\alpha_3$ such that the overall T1E rate is $\alpha_\cap$}
we need the distribution of \soutr{the random variable}
$K_3 \given \{K_{2} > a_{2}\}$ \hl{ where $K_n$ is defined in \eqref{eq:TP}}.
The condition $K_2 > a_2$
is needed, because the assessment of success after $3$
studies requires that there was no success after $2$ studies.
Due to \eqref{eq:Pseq}, the density
$f(k_3 \given \{K_{2} > a_{2}\})$ can be obtained by a convolution
of the density of $K_{2} \given \{K_{2} > a_{2}\}$ and the
density of $-2 \log(1-p_3) \sim \Ga(1, 1/2)$, a gamma distribution with shape and rate parameters 1 and $1/2$.  The density of
$K_{2} \given \{K_{2} > a_{2}\}$  is a $\Ga(2, 1/2)$ density truncated to $K_{2} > a_{2}$,
which occurs with probability
$1 - \Pr(K_{2} \leq a_{2}) = 1 - \alpha_2$: 
\[
  f(k_{2} \given \{K_{2} > a_{2}\}) = \frac{1}{1 - \alpha_2}\Ga(k_{2}; 2, 1/2)
  {\mathbf 1}_{\{k_{2} > a_2\}},
\]
now $\Ga(x; a, b)$ denotes the density of the
gamma distribution at $x$ with shape parameter $a$ and
rate parameter $b$.

The
convolution of $K_{2} \given \{K_{2} > a_{2}\}$ and $-2 \log(1-p_3)$ therefore has density
\[
  f(k_3 \given \{K_{2} > a_{2}\}) = \left\{
    \begin{array}{ll}\frac{1}{1-\alpha_2} \int\limits_{a_2}^{k_3}
      \Ga(x; 2, 1/2) \Ga(k_3-x; 1, 1/2) dx & \mbox{ if } k_3 > a_{2} \\
      0 & \mbox{ else.} \end{array} \right. 
\]
Numerical integration of $f(k_3 \given \{K_{2} > a_{2}\})$ can be
used to compute the cumulative distribution function
\begin{equation}\label{eq:Fvn}
  F(k_3 \given \{K_{2} > a_{2}\}) = \Pr( K_3 \leq k_3 \given \{ K_{2} > a_{2} \}).
\end{equation}
Root-finding methods are then used to determine the \soutr{value}\hl{adjusted
  level $\alpha_3$ with budget $\chi^2_6(\alpha_3)$}
\soutr{where} \hl{such that}
$F( a_3 \given \{ K_{2} > a_{2} \}) =
\alphaoverall - \alpha_2 = (1-q) \cdot \alphaoverall$ holds.  \soutr{The corresponding value of the nominal level $\alpha_3$ can finally be obtained with the
quantile function of the gamma distribution.}

\subsection{Edgington's method}
\hl{The adjusted level $\alpha_2 = q \cdot \alpha_\cap$ with budget
$b_2=\sqrt{2 \alpha_2}$ is chosen to spend the proportion $q$ of $\alpha_\cap$ on $\alpha_2$. }
\soutr{Let $b_k$ denote the available budget based on the adjusted significance level $\alpha_k$, $k=2,3$.}
To compute \hl{the adjusted level $\alpha_3$ such that the overall T1E rate is $\alpha_\cap$} \soutr{the T2E rate of the sequentzial Edgington method,} we need the distribution of \soutr{the random variable}
$E_3 \given \{E_{2} > b_{2}\}$ \hl{ where $E_n$ is defined in \eqref{eq:En}}.
Due to \eqref{eq:Eseq}, the density
$f(e_3 \given \{E_{2} > b_{2}\})$ can be obtained by a convolution
of the density of $E_{2} \given \{E_{2} > b_{2}\}$ and the
density of $p_3 \sim \Uni(0, 1)$.  The density of the first
term is a Irwin-Hall density truncated to $E_{2} > b_{2}$,
which occurs with probability
$1 - \Pr(E_{2} \leq b_{2}) = 1 - \alpha_{2}$:
\[
  f(e_{2} \given \{E_{2} > b_{2}\}) = \frac{1}{1 - \alpha_{2}}\IH(e_{2}; n=2)
  {\mathbf 1}_{\{e_{2} > b_2\}},
\]
here $\IH(x; n)$ denotes the density of the
Irwin-Hall distribution with parameter $n$. For $n=2$ this is
\begin{equation}\label{eq:IHpdf}
\IH(x; n=2) = \left\{\begin{array}{rl}
                          x & \mbox{ for } 0 \leq x < 1, \\
                          2-x & \mbox{ for } 1 \leq x \leq 2.
  \end{array}\right.
\end{equation}
The
convolution of $E_{2} \given \{E_{2} > b_{2}\}$ and $p_3$ therefore has density
\[
  f(e_3 \given \{E_{2} > b_{2}\}) = \left\{
    \begin{array}{ll}\frac{1}{1- \alpha_2} \int\limits_{b_2}^{e_3}
      \IH(x; 2) dx & \mbox{ if } e_3 > b_{2} \\
      0 & \mbox{ else.} \end{array} \right. 
    \]
Note that the integral $\int\limits_{b_2}^{e_3}
      \IH(x; 2) dx = \int\limits_{0}^{e_3}
      \IH(x; 2) dx - \int\limits_{0}^{b_2}
      \IH(x; 2) dx$ can be easily calculated based on the cdf of the $\IH(2)$ distribution, which is analytically available from \eqref{eq:IHpdf}. 
    Numerical integration of $f(e_3 \given \{E_{2} > b_{2}\})$ finally gives the cumulative distribution function
\begin{equation}\label{eq:Fvn}
  F(e_3 \given \{E_{2} > b_{2}\}) = \Pr( E_3 \leq e_3 \given \{ E_{2} > b_{2} \}).
\end{equation}
Root-finding methods are then used to determine the \soutr{value} \hl{adjusted
  level $\alpha_3$ with budget $b_3=\sqrt[3]{6 \alpha_3}$}
\soutr{where} \hl{such that}
$F( b_3 \given \{ E_{2} > b_{2} \}) =
\alphaoverall - \alpha_2 = (1-q) \cdot \alpha_\cap$ holds.
\soutr{The corresponding \soutr{value of
the nominal} \hl{adjusted} level $\alpha_3$ can finally be obtained with the quantile function of the Irwin-Hall distribution. }

\subsection{Held's method}
\hl{The adjusted level $\alpha_2 = q \cdot \alpha_\cap$ with budget
$c_2=4/\left[\Phi^{-1}(1-2 \alpha_2) \right]^2$ is chosen to spend the proportion $q$ of $\alpha_\cap$ on $\alpha_2$. }  \soutr{Let $c_k$ denote the budget based on the adjusted significance level $\alpha_k$, $k=2,3$.}
To compute \hl{the adjusted level $\alpha_3$ such that the overall T1E rate is $\alpha_\cap$} \soutr{the T1E rate of Held's method} we need the distribution of \soutr{the random variable}
$H_3 \given \{H_{2} > c_{2}\}$ \hl{ where $H_n$ is defined in \eqref{eq:Hn}}.
\soutr{Ignore the signs of $Z_i$, $i=1,2,3$, for the moment. }
Due to \eqref{eq:Hseq}, the density
$f(h_3 \given \{H_{2} > c_{2}\})$ can be obtained by a convolution
of the density of $H_{2} \given \{H_{2} > c_{2}\}$ and the
density of $1/Z_3^2 \sim \IG(1/2, 1/2)$\hl{, an inverse gamma distribution}.  The density of the first
term is an $\IG(1/2, 2)$ density truncated to $H_{2} > c_{2}$,
which occurs with probability
$1 - \Pr(H_{2} \leq c_{2}) = 1 - 4 \, \alpha_2$ \soutr{(ignoring the signs of $Z_1$ and $Z_2$)} \hl{(the factor 4 enters here because $H_2$ doesn't take the signs of $Z_1$ and $Z_2$ into account)}:
\[
  f(h_{2} \given \{H_{2} > c_{2}\}) = \frac{1}{1 - 4 \, \alpha_2}\IG(h_{2}; 1/2, 2)
  {\mathbf 1}_{\{h_{2} > c_2\}},
\]
now $\IG(x; a, b)$ denotes the density of the
inverse gamma distribution at $x$ with shape parameter $a$ and
rate parameter $b$:
\[
  \IG(x; a, b) = \left\{\begin{array}{rl}
                          \frac{b^{a}}{\Gamma(a)}x^{-(a+1)}\exp(-b/x) & \mbox{ for } x>0 \\
                        0 & \mbox{ else. } \end{array}\right.
\]

The
convolution of $H_{2} \given \{H_{2} > c_{2}\}$ and $1/Z_3^2$ therefore has density
\[
  f(h_3 \given \{H_{2} > c_{2}\}) = \left\{
    \begin{array}{ll}\frac{1}{1-4 \alpha_2} \int\limits_{c_2}^{h_3}
      \IG(x; 1/2, 2) \IG(h_3-x; 1/2, 1/2) dx & \mbox{ if } h_3 > c_{2} \\
      0 & \mbox{ else.} \end{array} \right. 
\]
Numerical integration of $f(h_3 \given \{H_{2} > c_{2}\})$ can be
used to compute the cumulative distribution function
\begin{equation}\label{eq:Fvn}
  F(h_3 \given \{H_{2} > c_{2}\}) = \Pr( H_3 \leq h_3 \given \{ H_{2} > c_{2} \}).
\end{equation}
Root-finding methods are then used to determine the \soutr{value} \hl{adjusted
  level $\alpha_3$ with budget $c_3=9/\left[\Phi^{-1}(1-4 \alpha_3) \right]^2$}
\soutr{where} \hl{such that}
$F( c_3 \given \{ H_{2} > c_{2} \})\hl{/8} =
\alphaoverall - \alpha_2 = (1-q) \cdot \alpha_\cap$ holds.   \hl{Division by $2^3=8$ is required
  because $H_3$ doesn't take the signs of $Z_1$, $Z_2$ and $Z_3$ into account.} \soutr{The corresponding {value of
the nominal} level $\alpha_3$ can be finally obtained from
\eqref{eq:alphaH.to.cH} and \eqref{eq:Hn} with $n=3$ and $\alphaoverall$ replaced by $\alpha_3 =
\left[1-\Phi(3/\sqrt{c_3}) \right]/4$.}

\section{Stopping Probabilities}\label{app:stopping}

\begin{figure}
\begin{knitrout}
\definecolor{shadecolor}{rgb}{0.969, 0.969, 0.969}\color{fgcolor}
\includegraphics[width=\maxwidth]{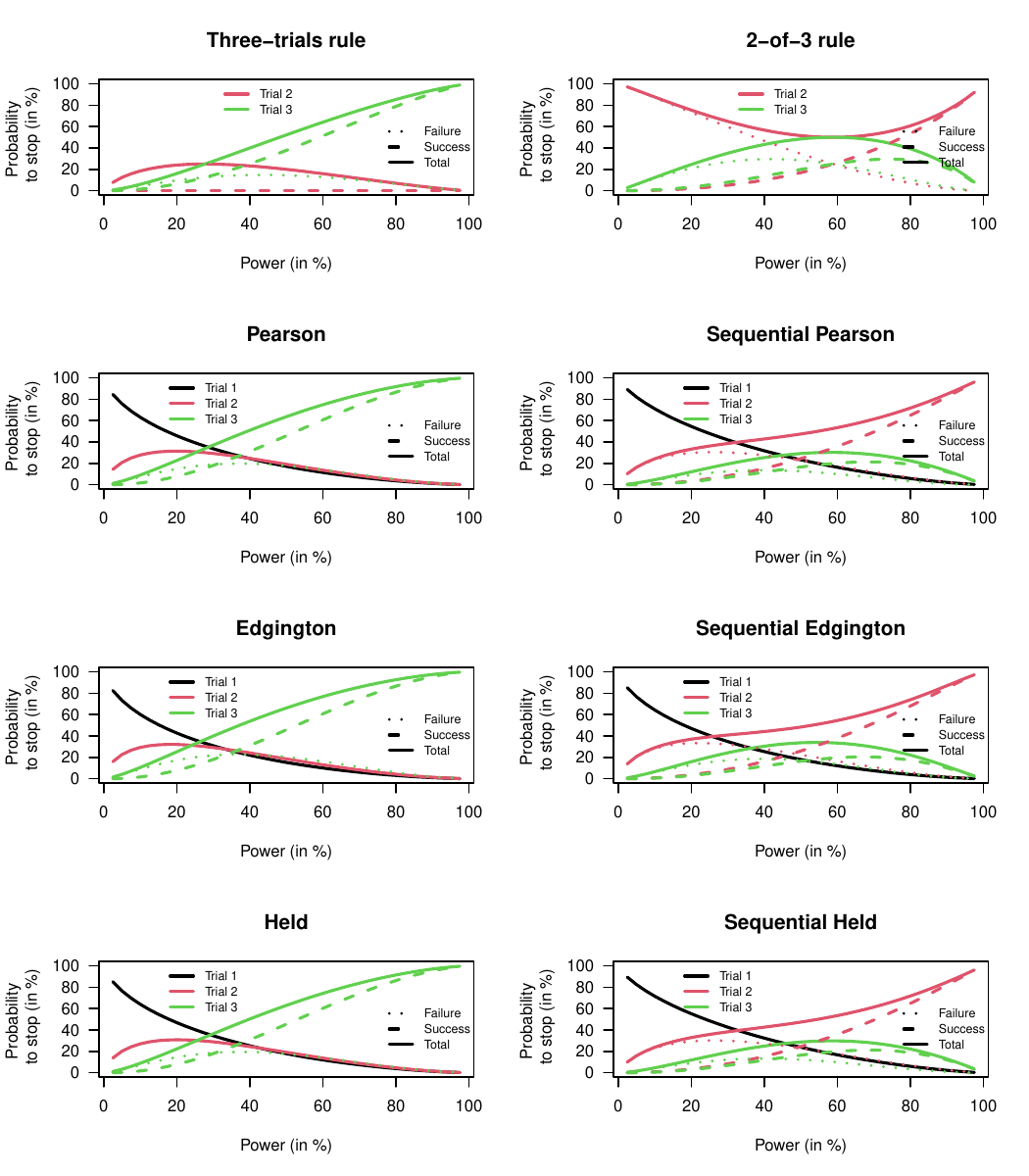} 
\end{knitrout}
\caption{Probabilities to stop after the different trials. Dotted
  lines represent stopping for failure. Dashed lines represent
  stopping for success. Solid lines represent stopping for either failure or success.
  \label{fig:Stop}}
\end{figure}

Figure \ref{fig:Stop} shows for the different approaches the probability to stop after the first,
second or third trial as a function of the power of the individual
trials varying between 2.5
and 97.5\%. 
The three-trials and 2-of-3 rule offers no possibility to stop after the first trial,
so only two lines are shown.

\section{\hl{Sample size reduction of Edgington's method}}
\label{app:SSReduction}
\begin{figure}
\begin{knitrout}
\definecolor{shadecolor}{rgb}{0.969, 0.969, 0.969}\color{fgcolor}
\includegraphics[width=\maxwidth]{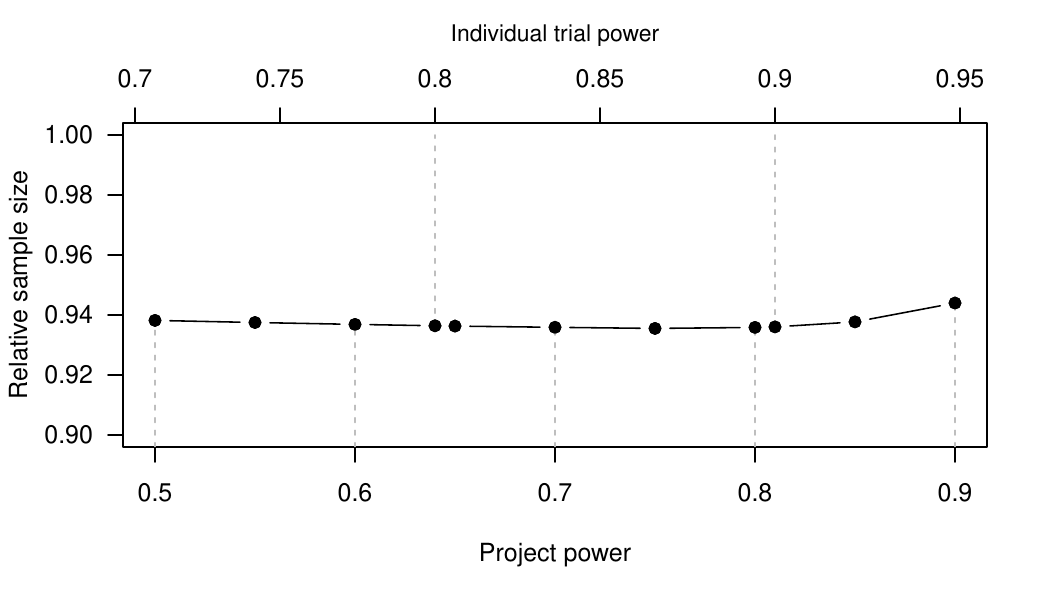} 
\end{knitrout}
\caption{\label{fig:SSReduction} \hl{Relative sample size required to achieve the same project power as the two-trials rule with Edgington's method. }}
\end{figure}

\hl{Suppose two trials are powered at significance level $\alpha$ with
individual trial power $1-\beta$.  The distribution of the the sum of the two
$p$-values can then be derived as the convolution of the distribution of
the two $p$-values under the alternative \cite{Hung_etal1997}, which
can be used to compute the project power of Edgington's method
numerically. Root-finding methods can then be applied to find the
power $1-\beta'$ that achieves the project power $(1-\beta)^2$ of the
two-trials rule. The relative sample size then is
\[
\frac{n'}{n} = \left\{
\frac{\Phi^{-1}(1-\alpha) + \Phi^{-1}(1-\beta')}{\Phi^{-1}(1-\alpha) + \Phi^{-1}(1-\beta)}\right\}^2
\]
and the sample size reduction is $1-n'/n$.
Figure \ref{fig:SSReduction} shows that the reduction is fairly constant around 6\%
for a wide range of values of the project power $(1-\beta)^2$.}

\end{appendix}
\end{document}